\documentclass[12pt]{article}
\usepackage{makeidx}
\usepackage{amsmath}
\usepackage{amsfonts}
\usepackage{amssymb}
\usepackage{graphicx}
\usepackage{cite}
\usepackage{xcolor}
\usepackage{amsthm}

\makeatletter
\g@addto@macro\th@plain{\thm@headpunct{:}}
\makeatother
\theoremstyle{plain}
\newtheorem*{acknowledgement*}{Acknowledgement}
\input epsf.sty
\newtheorem{theorem}{Theorem}
\newtheorem{acknowledgement}[theorem]{Acknowledgement}

\makeatletter \@addtoreset{equation}{section}
\renewcommand{\theequation}{\thesection.\arabic{equation}}
\input{tcilatex}
\begin{document}

\title{\textbf{Quantum line operators from Lax pairs}}
\author{El Hassan Saidi \\
{\small 1. LPHE-MS, Science Faculty}, {\small Mohammed V University in
Rabat, Morocco}\\
{\small 2. Centre of Physics and Mathematics, CPM- Morocco}}
\maketitle

\begin{abstract}
Motivated by the realisation of Yang-Baxter equation of 2d Integrable models
in the 4d gauge theory of Costello-Witten-Yamazaki (CWY), we study the
embedding of integrable 2d Toda field models inside this construction. This
is done by using the Lax formulation of 2d integrable systems and by
thinking of the standard Lax pair $L_{\pm }$ in terms of components of CWY
gauge connection propagating along particular directions in the gauge
bundle. We also use results of the CWY theory to build quantum line
operators for 2d Toda models and compute the one loop contribution of two
intersecting lines exchanging one gluon. Other features like local
symmetries and comments on extension of our method to other 2d integrable
models are also discussed. We also comment on\textrm{\ }some basic points
that still need a refinement before talking about a fully consistent
embedding of Lax pairs into CWY theory.\newline
\end{abstract}

\section{Introduction}

Yang-Baxter equation (YBE) generally expressed $%
R_{12}R_{13}R_{23}=R_{23}R_{13}R_{12}$ is a basic hyper- matrix relation of
2d quantum integrable models that has been subject to many studies \cite%
{1A,2A,3A} and has been revealed to be important for several issues; for
example in the formulation of Hopf algebras and quantum groups \cite%
{1B,2B,3B,4B,6B}, and in dealing with knots of 3d Chern-Simons gauge theory 
\cite{1C,1CA,1CB,1CC}; as well as with relationships concerning integrable
lattice models from the view of topological quantum field theories \cite%
{2C,3C}; see also \cite{4C} for a Gauge/YBE correspondence linking SU$\left(
N\right) $ quiver gauge theories with the partition function of 2d
integrable spin models. This non linear equation in R-matrix has initially
appeared in two different contexts of integrable models as a sufficient
condition for exact solvability. It appeared first in the factorisation
property of many body scattering amplitudes of relativistic QFT \cite{1D,2D}%
; and second for the transfer matrix of statistical models to commute for
different values of the spectral parameters \cite{3D,4D}.

\  \  \  \  \newline
Recently a formal topological four-dimensional gauge theory has been
constructed in \cite{1F} to deal with the fundamentals of the Yang-Baxter
equation in terms of a non abelian gauge potential $\mathcal{A}_{\mu }$ and
of gauge invariant quantum line operators $W_{\varrho _{i}}\left(
K_{i}\right) $ as basic quantities that are behind the derivation of the
solutions of the matrix R and also behind the study of its quantum
properties. Based on previous results of \cite{2F,3F} and nicely motivated
in \cite{1G}, this construction uses the power of the quantum field theory
(QFT) method \textrm{to study quantum properties of intersecting multi- line
configurations }$K_{1},...K_{n}$ supporting non local observables $%
W_{\varrho _{1}}\left( K_{1}\right) ...,W_{\varrho _{n}}\left( K_{n}\right) $%
. It has allowed to rederive known results on 2d integrable systems in a
nice manner; and has permitted moreover to obtain new involved ones like the
RTT presentation for Yangians $\mathcal{Y}\left( \mathcal{G}_{c}\right) $ 
\cite{1J}. For other applications, see also \cite{4F} dealing with
unification of integrability in supersymmetric models and for the six
dimensional origin of topological invariant constructions. The formal gauge
field theory of \cite{1F} to which we refer hereafter to as the CWY theory
--- CWY for Costello-Witten-Yamazaki --- is a topological gauge theory with
1-form gauge connection $\mathcal{A}=\mathcal{A}_{\mu }dX^{\mu }$ described
by a partial gauge potential $\mathcal{A}_{\mu }=\left( \mathcal{A}_{x},%
\mathcal{A}_{y},\mathcal{A}_{\bar{\zeta}}\right) $ living on 4d manifolds $%
\mathbb{M}_{4}$ that factorises as the product of two Riemann surfaces $%
\Sigma $ and $\mathcal{C}$. In the CWY approach, the well known three kinds
of quasi- classical solutions of the Yang-Baxter equation (rational,
trigonometric, elliptic) and their underlying quantum group symmetries \cite%
{1H} have been derived from specific aspects of the 4d space $\mathbb{M}%
_{4}=\Sigma \times \mathcal{C}$, hosting the CWY gauge theory, and from
properties the quantum line operators like framing anomaly, fusion of lines
due to scaling symmetry as well as line operator product expansions \cite{1J}%
.

\  \  \  \  \  \newline
In this paper, we contribute to this matter by studying the embedding of a
family of 2d integrable QFT models inside the CWY theory and use this
approach to get more insight on properties of their quantum integrability.
Concretely, we consider conformal Toda field theory in 2d, which constitute
a class of 2d integrable QFT models based on finite dimensional Lie
algebras, and study its embedding into the 4d CWY theory. The leading 2d QFT
model in this finite Toda QFT$_{2}$ family is given by the well known
integrable Liouville theory which is associated with sl$\left( 2\right) $
algebra and which we take it here as an illustrating example. By focussing
on this leading sl$\left( 2\right) $ based model, we show that the insertion
of the Liouville field into CWY can be done by using the Lax formalism
allowing to linearise the Liouville equation by help of a pair of operators $%
L_{\pm }$. Here, the Lax pair $\left( L_{+},L_{-}\right) $ is thought of as
given by a particular gauge field configuration that solves the field
equation of motion of the CWY vector potential. Using this approach, we
develop a method to build quantum line operators $W\left( K_{i}\right) $ for
Liouville theory that define non local observables of the theory and which
give the bridge between Liouville field and the Yang-Baxter equation of 2d
integrable models. As an application of the construction, we calculate as
well the amplitude at one loop order of two intersecting lines exchanging
one gluon. We also comment on some basic points that still need a refinement
before talking about a fully consistent embedding of Lax pair formalism into
CWY theory.

\  \  \  \  \  \newline
The organisation of this paper is as follows: In section 2, we review some
useful aspects on the 4d CWY theory. In section 3, we recall the classical
Liouville equation and give a list of some of its remarkable properties
which are relevant to the present study. In section 4, we develop the 2d Lax
formalism for Liouville field and study links with the CWY connection. In
section 5, we study the embedding of the Liouville equation into the CWY
modeling and build the associated quantum lines. In section 6, we compute
the amplitude of two intersecting lines exchanging one gluon. Section 7 is
devoted to the conclusion and to comments while section 8 is devoted to an
appendix where results for Toda fields are reported.

\section{CWY theory: an overview}

\label{sec2} In this section, we give a brief review of those tools of the
Costello-Witten-Yamazaki gauge theory that are useful for dealing with the
modeling of the solutions of the Yang-Baxter equation and for the study of
interacting line operators. Some of these tools will be rephrased so that
they can be used in next sections when considering the application of the
CWY theory to approach finite Toda QFT$_{2}$'s; in particular the integrable
2d Liouville theory and the building of its quantum line operators.

\subsection{Formal field action}

\begin{figure}[tbph]
\begin{center}
\hspace{0cm} \includegraphics[width=10cm]{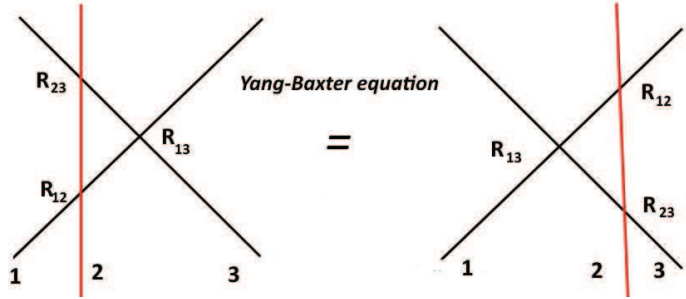}
\end{center}
\par
\vspace{-0.5 cm}
\caption{The relation --- $R_{12}R_{13}R_{23}=R_{23}R_{13}R_{12}$ ---
expressed by a graphic representation of Yang-Baxter equation of 2d
integrable systems; the three lines characterised by spectral parameters
"complex rapidities" $\protect \zeta _{1},\protect \zeta _{2},\protect \zeta %
_{3}$ cross each others. The regularity of the displacement of the red line
from left to right of the crossing is ensured by Diff$\left( \Sigma \right) $%
. }
\label{ybe}
\end{figure}

\begin{figure}[tbph]
\begin{center}
\hspace{0cm} \includegraphics[width=10cm]{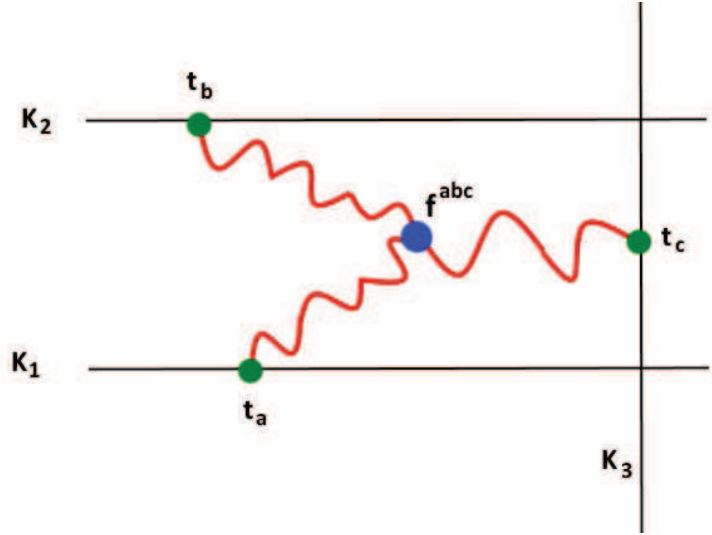}
\end{center}
\par
\vspace{-0.5 cm}
\caption{An example of Feynman diagram with three interacting line operators 
$W_{\protect \varrho _{1}}\left( K_{1}\right) ,W_{\protect \varrho _{2}}\left(
K_{2}\right) $ and $W_{\protect \varrho _{3}}\left( K_{3}\right) $ with
spectral parameters $\protect \zeta _{1},\protect \zeta _{2}$ and $\protect%
\zeta _{3}$. Two line operators are taken as parallel in the topological
plane with equations $y=0$ and $y=\protect \epsilon $. The third is given by $%
x=0$. The three lines interact through gluons with bulk interaction located
at $\left( x,y,\bar{\protect \zeta},\protect \zeta \right) .$}
\label{02}
\end{figure}

A manner to introduce the 4d CWY theory modeling the solutions of Yang-
Baxter (YB) equation $R_{12}R_{13}R_{23}=R_{23}R_{13}R_{12}$ in terms of
partial gauge fields $\mathcal{A}_{\mu }\left( X\right) $ is to start from
the explicit expression of the field action $\mathcal{S}_{cwy}\left[ 
\mathcal{A}\right] $ and the gauge invariant observables $W_{\varrho _{i}}%
\left[ K_{\zeta _{i}}\right] $ of the theory. Then, use the path integral
method and Feynman diagram rules to approach the quasi-classical solutions
of the R- matrices $R_{ij}\left( \zeta _{ij}\right) $ by using crossing
quantum line operators $W_{\varrho _{i}}\left[ K_{\zeta _{i}}\right] $
exchanging gauge particles as in figure \ref{02}.

\  \  \  \  \newline
The action $\mathcal{S}_{cwy}$ is a \emph{formal} 4d functional living on a
4 space $\mathbb{M}_{4}$ given by the cross product of two Riemann surfaces; 
$\mathbb{M}_{4}=\Sigma \times \mathcal{C}$. It reads in terms of
differential forms as%
\begin{equation}
\mathcal{S}_{cyw}\left[ \mathcal{A},\omega \right] =\frac{1}{2\pi }%
\int_{\Sigma \times \mathcal{C}}\boldsymbol{L}_{4}\left( \mathcal{A},\omega
\right)  \label{1}
\end{equation}%
where the 4-form Lagrangian $\boldsymbol{L}_{4}$ has some special features
of which the two following ones: $\left( i\right) $ it is holomorphic in the
partial gauge connection $\mathcal{A}$ valued in a complexified finite
dimensional Lie algebra $\mathcal{G}_{c}$; i.e: no adjoint conjugate $%
\mathcal{A}^{\dagger }\equiv \overline{\mathcal{A}}$, and $\left( ii\right) $
it is given by the exterior product on $\Sigma \times \mathcal{C}$, 
\begin{equation}
\boldsymbol{L}_{4}=\mathbf{\omega }_{1}\wedge \boldsymbol{\Omega }_{3}
\end{equation}%
with 3-form $\boldsymbol{\Omega }_{3}$ of Chern-Simons (CS) type 
\begin{equation}
\boldsymbol{\Omega }_{3}=\mathcal{A}\wedge d\mathcal{A}+\frac{2}{3}\mathcal{A%
}\wedge \mathcal{A}\wedge \mathcal{A}  \label{m}
\end{equation}%
and $\mathbf{\omega }_{1}=\omega _{\zeta }d\zeta $ a holomorphic 1-form
living on the complex curve $\mathcal{C}$ with no zeros; but may have poles.
The two real surfaces $\Sigma $ and $\mathcal{C}$ play an important role in
the CWY construction; the $\Sigma $ hosts the real curves $K_{\zeta }$
supporting the line operators $W_{\varrho }\left[ K_{\zeta }\right] $; and
the $\mathcal{C}$ gives the coordinate $\zeta $ of the curve $K_{\zeta }$
inside $\mathbb{M}_{4}$ and then the spectral parameter of the R-matrices.
By denoting the four local coordinates of $\mathbb{M}_{4}=\Sigma \times 
\mathcal{C}$ like 
\begin{equation}
X^{M}=\left( x,y;\bar{\zeta},\zeta \right)
\end{equation}%
with $\left( X_{1},X_{2}\right) =\left( x,y\right) $ for the Riemann surface 
$\Sigma $ and $\zeta =X_{3}+iX_{4}$ for the complex line $\mathcal{C}$, we
have $\boldsymbol{\Omega }_{3}=\Omega _{\left[ \mu \nu \sigma \right]
}dX^{\mu }\wedge dX^{\nu }\wedge dX^{\sigma }$ with%
\begin{equation}
\Omega _{\left[ \mu \nu \sigma \right] }=\mathcal{A}_{[\mu }\partial _{\nu }%
\mathcal{A}_{\sigma ]}+\frac{2}{3}\mathcal{A}_{[\mu }\mathcal{A}_{\nu }%
\mathcal{A}_{\sigma ]}
\end{equation}%
and where $\mathcal{A}_{\mu }=t_{a}\mathcal{A}_{\mu }^{a}$ with $t_{a}$
standing for a basis of generators of $\mathcal{G}_{c}$. For the holomorphic
1-form, we have the three following possibilities depending on the nature of
the complex curve $\mathcal{C}$,%
\begin{equation}
\begin{tabular}{|l|l|}
\hline
{\small \ complex curve} $\mathcal{C}$ \  \  & {\small \ holomorphic 1-form}$%
\  \mathbf{\omega }_{1}$ \\ \hline
$\  \  \  \  \  \  \  \  \  \  \mathbb{C}$ & $\  \  \  \  \  \  \  \  \  \ d\zeta $ \\ \hline
$\  \  \  \  \  \  \  \  \  \mathbb{C}^{\times }$ & $\  \  \  \  \  \ d\left( \log \zeta
\right) $ \\ \hline
$\  \  \  \mathbb{C}/\left( \mathbb{Z+\tau Z}\right) $ \  & $\  \  \  \  \  \  \  \  \
\  \  \ d\zeta $ \\ \hline
\end{tabular}%
\end{equation}%
\begin{equation*}
\end{equation*}%
Notice that for convenience, we use below three kinds of space indices to
refer to the space coordinates and to the fields living on $\Sigma \times 
\mathcal{C}$. The capital latin index $M$ to refer to full space vectors
like $X^{M}=\left( x,y;\bar{\zeta},\zeta \right) $ or equivalently $\left(
X^{+},X^{-};\bar{\zeta},\zeta \right) $ with $X^{\pm }=x\pm iy$. The Greek
index $X^{\mu }$ to refer to the subspace $\left( X^{+},X^{-};\bar{\zeta}%
\right) $ where spread the 3-form $\boldsymbol{\Omega }_{3}$ directions.
Finally the tiny latin index like in $X^{m}$ to refer to $\left(
X^{+},X^{-}\right) $ and in general for vectors on $\Sigma $. So, we have
the convention notation%
\begin{equation}
M=\mu ;\zeta \qquad ,\qquad \mu =m;\bar{\zeta}\qquad ,\qquad m=+,-
\end{equation}%
The field equation of the \emph{partial} 1-form gauge potential $\mathcal{A}=%
\mathcal{A}_{\mu }dX^{\mu }$ is obtained by the functional variation of $%
\mathcal{S}_{cyw}$; it is given, in differential form language, by 
\begin{equation}
\frac{\delta \mathcal{S}_{cyw}}{\delta \mathcal{A}}=0\qquad \Rightarrow
\qquad \mathbf{\omega }_{1}\wedge \mathcal{F}=0  \label{fe}
\end{equation}%
\ with non vanishing holomorphic 1-form, $\mathbf{\omega }_{1}\mathbf{\neq 0}
$; and where the 2-form $\mathcal{F}=\mathcal{F}_{\mu \nu }dX^{\mu }\wedge
dX^{\nu }$ is nothing but the gauge curvature of the partial vector
potential $\mathcal{A}$ filling the $\left( x,y;\bar{\zeta}\right) $ space
directions and reading as follows%
\begin{equation}
\mathcal{F}=d\mathcal{A}+\mathcal{A}\wedge \mathcal{A}  \label{ef}
\end{equation}%
The gauge field equation of motion (\ref{fe}) is naturally solved as $%
\mathcal{F}=0$ requiring $\mathcal{F}_{\mu \nu }=0$ and which reads in terms
of the $\mathcal{A}_{\mu }$ components of the partial gauge connection as
follows%
\begin{eqnarray}
\partial _{+}\mathcal{A}_{-}-\partial _{-}\mathcal{A}_{+}+\left[ \mathcal{A}%
_{+},\mathcal{A}_{-}\right] &=&0  \notag \\
\partial _{\bar{\zeta}}\mathcal{A}_{+}-\partial _{+}\mathcal{A}_{\bar{\zeta}%
}+\left[ \mathcal{A}_{\bar{\zeta}},\mathcal{A}_{+}\right] &=&0  \label{29} \\
\partial _{\bar{\zeta}}\mathcal{A}_{-}-\partial _{-}\mathcal{A}_{\bar{\zeta}%
}+\left[ \mathcal{A}_{\bar{\zeta}},\mathcal{A}_{-}\right] &=&0  \notag
\end{eqnarray}%
The $\mathcal{A}_{\mu }$ defines a flat gauge bundle on $\Sigma $ which
varies holomorphically as we move on $\mathcal{C}$. In \textrm{section 5},
we will study a particular solution of these component field equations;
before that, notice the three useful features. First, eqs (\ref{1}) and (\ref%
{fe}) have rich local symmetries allowing to perform several operations like
for instance doing gauge transformations or also moving in a safe way line
operators from the left to the right in a system with apparently crossing
lines configurations as schematized by figure (\ref{ybe}). 
In addition to the complexified gauge symmetry with Lie algebra $\mathcal{G}%
_{c}$ allowing freedom in changing the vector potential as $\mathcal{A}_{\mu
}^{\prime }=g^{-1}\mathcal{A}_{\mu }g+g^{-1}\partial _{\mu }g$, the action $%
\mathcal{S}_{cyw}$ and the field equation are also invariant under $%
diff\left( \Sigma \right) ,$ the group of diffeomorphisms of $\Sigma $ given
by general coordinates transformations $X^{\prime m}=f^{m}\left(
X^{+},X^{-}\right) $; and are invariant as well under $Hol\left( \mathcal{C}%
\right) ,$ the group of holomorphic transformations on $\mathcal{C}$ with
local coordinate $\zeta $. Second, from $\mathcal{S}_{cyw}\left[ \mathcal{A}%
,\omega \right] $, one can determine the free gauge propagators 
\begin{equation}
G_{\mu \nu }^{ab}\left( X-X^{\prime }\right) =\left \langle \mathcal{A}_{\mu
}^{a}\left( X\right) \mathcal{A}_{\nu }^{b}\left( X^{\prime }\right) \right
\rangle
\end{equation}%
represented by the wavy red line in figure \ref{02}. One can also learn from
the interacting part of the action the structure of the 3-vertex $\Gamma
_{\mu \nu \sigma }^{abc}$ of the tri-vector fields coupling $\left \langle 
\mathcal{A}_{\mu }^{a}\mathcal{A}_{\nu }^{b}\mathcal{A}_{\sigma
}^{c}\right
\rangle $. Using p-form language by killing the $\mu $- space
indices with the help of the differentials $dX^{\mu }$, the free propagators 
$G_{\mu \nu }^{ab}$ get mapped to 2-form propagators 
\begin{equation}
\mathcal{P}^{ab}=\delta ^{ab}\mathcal{P}  \label{pp}
\end{equation}%
with $\mathcal{P}=G_{\mu \nu }dX^{\mu }\wedge dX^{\nu }$ and where 
\begin{equation}
G_{\mu \nu }=\frac{1}{4\pi }\varepsilon _{\mu \nu \sigma }\eta ^{\sigma \tau
}\frac{\partial }{\partial X^{\tau }}\left( \frac{1}{\left( x-x^{\prime
}\right) ^{2}+\left( y-y^{\prime }\right) ^{2}+\left \vert \zeta -\zeta
^{\prime }\right \vert ^{2}}\right)  \label{pro}
\end{equation}%
Similarly, the vertex of the coupling of the three gauge fields carries only
adjoint representation group indices and reads in terms of the $f^{abc}$
structure constant of $\mathcal{G}_{c}$ as follows 
\begin{equation}
\Gamma ^{abc}=\frac{i}{2\pi }f^{abc}d\zeta  \label{ve}
\end{equation}%
The third feature we want to comment on here concerns observables $\mathcal{O%
}$ in the CWY theory and their quantum properties. In this regards, it is
interesting to notice that the on shell vanishing of the gauge curvature $%
\mathcal{F}$ teaches us that the CWY construction behaves somehow like the
3d topological CS gauge theory in the sense that there are no local
observables $\mathcal{O}\left( \mathcal{F}\right) $ in the CWY theory that
can built by using gauge invariant polynomials in the curvature $\mathcal{F}$%
; they vanish identically due to gauge invariance and the field equation of
motion. However, the construction of non trivial observables in the CWY
theory is still possible; it needs considering other gauge invariants that
are non local quantities as described in what follows for the example $%
\mathbb{M}_{4}=\mathbb{R}^{2}\times \mathbb{C}$. 

\subsection{Observables}

Despite the on shell vanishing of the 2-form gauge curvature $\mathcal{F}=0$%
, we can still construct observables $\mathcal{O}$ in the 4d gauge theory of
Costello-Witten-Yamazaki; they are given by non local invariant operators,
some of them will be described in a moment. The vacuum expectation values
(VEV) of these observable are given by the usual path integral formulation 
\begin{equation}
\left \langle \mathcal{O}\right \rangle =\frac{\dint \left[ DA\right] 
\mathcal{O}\exp \left( \frac{i}{\hbar }\mathcal{S}_{cyw}\right) }{\dint %
\left[ DA\right] \exp \left( \frac{i}{\hbar }\mathcal{S}_{cyw}\right) }
\end{equation}%
with $\mathcal{S}_{cyw}=\frac{1}{2\pi }\int_{\mathbb{R}^{2}\times \mathbb{C}%
}d\zeta \wedge \Omega _{3}$; and where $\hbar $ is a\textrm{\ }parameter
scaling as length with powers capturing the quantum loop corrections. A
particular class of these gauge invariant quantities is given by the line
operators 
\begin{equation}
W_{\varrho }\left[ \mathbf{\varphi }\left( K_{\zeta }\right) \right]
=Tr_{\varrho }\left[ P\exp \left( \doint_{K_{\zeta }}\mathcal{A}\right) %
\right]  \label{wr}
\end{equation}%
with $P$ referring to the path ordering and $\varrho $ to some
representation of the finite dimensional $\mathcal{G}_{c}$. The presence of
the trace is to ensure invariance under gauge transformations. Like for
Wilson line operators of the 3d Chern-Simons gauge theory, these gauge
invariant quantities of the CWY theory are based as well on the holonomy
term of the 1-form gauge potential $\mathcal{A}$; but along particular loops 
$K_{_{\zeta }}$ in the 4d space 
\begin{equation}
\mathbf{\varphi }\left( K_{\zeta }\right) =\doint \nolimits_{K_{_{\zeta }}}%
\mathcal{A}
\end{equation}%
Indeed, the real curves $K_{_{\zeta }}$ involved in the building of the $%
W_{\varrho }\left[ \mathbf{\varphi }\left( K_{\zeta }\right) \right] $
operators are very special in the sense that they should belong to the
surface $\Sigma $ part of $\mathbb{M}_{4}$; but also live at some point $%
\zeta $ in $\mathcal{C}$. These $K_{_{\zeta }}$'s are then described by real
algebraic equations relating the $x,y$ variables like $y=f_{\zeta }\left(
x\right) $ where the complex $\zeta \in \mathcal{C}$ plays the role of a
spectral parameter. Therefore, the above holonomy should be treated as 
\begin{equation}
\mathbf{\varphi }\left( K_{\zeta }\right) =\doint_{K_{_{\zeta }}}t_{a}%
\mathcal{A}_{m}^{a}dX^{m}  \label{am}
\end{equation}%
with $X^{m}=\left( x,y\right) $ and $t_{a}$ the generators of $\mathcal{G}%
_{c}$. By using $X^{m}=\left( X^{+},X^{-}\right) $, we also have $\mathcal{A}%
=\mathcal{A}_{+}dX^{+}+\mathcal{A}_{-}dX^{-}$ on the curve $K_{_{\zeta }}$.
Notice that the gauge components $\mathcal{A}_{m}^{a}$ in above (\ref{am})
have a hidden structure due to the presence of the spectral parameter $\zeta 
$. Because of the coordinate dependence $\mathcal{A}_{m}^{a}=\mathcal{A}%
_{m}^{a}\left( x,y,\zeta ,\bar{\zeta}\right) $, one can define generalised
Wilson lines $\mathcal{W}_{\hat{\varrho}}\left( K_{\zeta }\right) $
extending (\ref{wr}). This is done by substituting the $\mathcal{A}_{\pm
}=t_{a}\mathcal{A}_{\pm }^{a}$ in (\ref{am}) by a holomorphic expansion in
the spectral parameter $\zeta $ like%
\begin{equation}
\boldsymbol{\hat{A}}_{\pm }\left( x,y,\zeta \right) :=\sum_{k=0}^{\infty
}t_{a,k}\hat{A}_{\pm }^{a\left( k\right) }\left( x,y\right)  \label{xp}
\end{equation}%
with 
\begin{equation}
t_{a,n}=t_{a}\otimes \zeta ^{n}
\end{equation}%
and where the $\hat{A}_{\pm }^{a\left( n\right) }\left( x,y\right) $ modes
in eq(\ref{xp}) may be imagined as given by the following modes in the $%
\zeta $- expansion where $\bar{\zeta}$ has been omitted, 
\begin{equation}
\hat{A}_{\pm }^{a\left( n\right) }\left( x,y\right) :=\frac{1}{n!}\left. 
\frac{\partial ^{n}}{\partial \zeta ^{n}}A_{\pm }^{a}\left( x,y;\zeta ,\bar{%
\zeta}\right) \right \vert _{\zeta =\bar{\zeta}=0}
\end{equation}%
Notice that the induced operators $t_{a,n}=t_{a}\otimes \zeta ^{n}$ generate
an infinite dimensional Lie algebra $\mathcal{G}\left[ \left[ \zeta \right] %
\right] $ containing the finite dimensional $\mathcal{G}_{c}$ as the
subalgebra of zero modes. In terms of the $\boldsymbol{\hat{A}}=\boldsymbol{%
\hat{A}}\left( x,y,\zeta \right) $, the generalised Wilson line operators
read therefore as follows 
\begin{equation}
\mathcal{W}_{\hat{\varrho}}\left( K_{\zeta }\right) =Tr_{\hat{\varrho}}\left[
P\exp \left( \dint \nolimits_{K_{\zeta }}\boldsymbol{\hat{A}}\right) \right]
\label{19}
\end{equation}%
where now $\hat{\varrho}$ stands for a representation of $\mathcal{G}\left[ %
\left[ \zeta \right] \right] $. Notice that the presence of the $Tr_{\hat{%
\varrho}}$ in defining above operator is somehow undesirable as it kills the
effect of $\mathcal{G}\left[ \left[ \zeta \right] \right] $ and makes the
generalisation meaningless; by following \textrm{\cite{1F}}, this difficulty
may be overcome by dropping the trace in defining $\mathcal{W}_{\hat{\varrho}%
}\left( K_{\zeta }\right) $; that is restricting (\ref{19}) to%
\begin{equation}
\mathcal{W}_{\hat{\varrho}}\left( K_{\zeta }\right) \text{ \ }\sim \text{ \ }%
Pe^{\mathbf{\hat{\varphi}}\left( K_{\zeta }\right) }\qquad ,\qquad \mathbf{%
\hat{\varphi}}\left( K_{\zeta }\right) =\dint \nolimits_{K_{\zeta }}%
\boldsymbol{\hat{A}}  \label{20}
\end{equation}%
Though apparently not invariant under gauge transformations $g=g\left(
X^{+},X^{-}\right) $ since the $e^{\mathbf{\hat{\varphi}}\left( K_{\zeta
}\right) }$ holonomy varies like $g^{-1}e^{\mathbf{\hat{\varphi}}\left(
K_{\zeta }\right) }g$; however this holonomy can be made gauge invariant if
considering the limit of lines $K_{\zeta }$ spreading\ to infinity in $%
\Sigma $ with the property $g\rightarrow I$ when $\left \vert X^{\pm
}\right
\vert \rightarrow \infty $. This restriction can be also justified
by the infrared-free limit of the CWY theory where the holonomy term behaves
as a gauge invariant quantity. \newline
With this brief review of CWY formalism and the rephrasing of some of its
tools, we come now to address the question on how to embed known 2d
integrable QFTs in the Costello-Witten-Yamazaki theory. In what follows, we
shall focuss on the explicit solutions of eqs(\ref{29}) by first considering
the restriction of these relations to the subspace $\Sigma $; so the three
relations reduces to the first one of (\ref{29}) and will be interpreted in
terms of 2d Lax equations of 2d integrable systems. After that, we turn to
study the solution of (\ref{29}) for the full $\Sigma \times \mathcal{C}$
with $\mathcal{G}_{c}$ a complex finite dimensional Lie algebra.

\section{Liouville equation and special aspects}

We begin by introducing briefly the classical Liouville equation in 2d space
by considering first both Lorentzian $\mathbb{R}^{1,1}$ and euclidian $%
\mathbb{R}^{2}$ signature; but focussing later on $\mathbb{R}^{2}$. We also
use this description to fix some convention notations. After that, we make
three comments on this 2d field equation which are helpful when studying the
embedding of Liouville field into CWY theory. Some aspects on finite 2d Toda
theory will be also commented.

\subsection{2d field action}

In real 2d space-time space $\mathbb{R}^{1,1}$ with 1+1 signature and local
coordinates $\mathrm{\rho }^{\mathrm{\alpha }}=\left( \sigma ,\tau \right) $%
, the classical Liouville equation is\ an integrable 2d field equation of
the form 
\begin{equation}
\frac{\partial ^{2}\phi }{\partial \tau ^{2}}-\frac{\partial ^{2}\phi }{%
\partial \sigma ^{2}}+\tilde{\kappa}e^{2\phi }=0  \label{s0}
\end{equation}%
with $\phi =\phi \left( \sigma ,\tau \right) $ a real 2d field and $\tilde{%
\kappa}$ a real constant parameter scaling as $\left( length\right) ^{-2}$.
This 2d field equation, which can be also presented like $\frac{\partial
^{2}\phi }{\partial \mathrm{\rho }^{+}\partial \mathrm{\rho }^{-}}+\kappa
e^{2\phi }=0$ with light cone coordinates $\mathrm{\rho }^{\pm }=\tau \pm
\sigma $ and $\frac{\partial }{\partial \mathrm{\rho }^{\pm }}=\frac{1}{2}(%
\frac{\partial }{\partial \tau }\pm \frac{\partial }{\partial \sigma })$,
can be derived from an action principle $\delta \mathcal{S}_{L}\left[ \phi %
\right] =0$ with 
\begin{equation}
\mathcal{S}_{L}\left[ \phi \right] \text{ \ }\sim \text{ \ }\int_{\Sigma
}\left( \frac{\partial \phi }{\partial \mathrm{\rho }^{-}}\frac{\partial
\phi }{\partial \mathrm{\rho }^{+}}-\frac{\kappa }{2}e^{2\phi }\right)
\label{s2}
\end{equation}%
where the real surface $\Sigma $ is given here by $\mathbb{R}^{1,1}$. This
action describes a non linear dynamics of the real 2d scalar field $\phi
=\phi \left( \mathrm{\rho }^{+},\mathrm{\rho }^{-}\right) $ to which we
refer below to as the classical Liouville field; the integral measure in (%
\ref{s2}) is given by $d\tau d\sigma =\frac{1}{2}d\mathrm{\rho }^{-}\wedge d%
\mathrm{\rho }^{+}$ with $\varepsilon _{-+}=\varepsilon ^{+-}=1$. The scalar
potential of the Liouville theory namely 
\begin{equation}
\mathcal{V}\left( \phi \right) =\frac{\kappa }{2}e^{2\phi }
\end{equation}%
has two special aspects: First, the factor $2$ in the argument of $e^{2\phi
} $ may be imagined in terms of the Cartan matrix $C_{11}=2$ of sl$\left(
2\right) $, it indicates how the Liouville theory can be extended to 2d Toda
theories\footnote{%
We refer to this class of 2d field models as finite Toda QFT$_{2}$; this
family is sometimes designated as conformal Toda QFT$_{2}$. Notice that
there exists also another class of Toda QFT$_{2}$ based on affine Lie
algebras and known as affine Toda theories \cite{1KA,2KA,3KA}.} based on
finite dimensional Lie algebra \cite{1K,2K,3K}. There, the 2d Toda fields $%
\left \{ \phi _{1},...,\phi _{r}\right \} $ extending the Liouville $\phi $
are given by r real scalars that can be also presented like 
\begin{equation}
\boldsymbol{\vec{\phi}}=\boldsymbol{\vec{\alpha}}_{1}\phi _{1}+\boldsymbol{%
\vec{\alpha}}_{2}\phi _{2}+...+\boldsymbol{\vec{\alpha}}_{r}\phi _{r}
\label{fi}
\end{equation}%
with $\boldsymbol{\vec{\alpha}}_{1},...,\boldsymbol{\vec{\alpha}}_{r}$
standing for the simple roots of $\mathcal{G}_{c}$. In affine Toda theories 
\cite{1KA,2KA,3KA}, eq(\ref{fi}) includes an extra time-like term $%
\boldsymbol{\vec{\alpha}}_{0}\phi _{0}$ with $\boldsymbol{\vec{\alpha}}_{0}$
the imaginary root of affine Lie algebras. For finite Toda QFT$_{2}$, the
field action reads as follows \cite{1KB,2KB,3KB},%
\begin{equation}
\mathcal{S}_{Toda}\left[ \phi _{1},...,\phi _{r}\right] \text{ \ }\sim \text{
\ }\int_{\Sigma }\left[ C_{ij}\frac{\partial \phi _{i}}{\partial \mathrm{%
\rho }^{-}}.\frac{\partial \phi _{j}}{\partial \mathrm{\rho }^{+}}%
-\sum_{i=1}^{r}\kappa _{i}e^{C_{ij}\phi _{j}}\right]  \label{tod}
\end{equation}%
with $C_{ij}=\frac{2}{\boldsymbol{\vec{\alpha}}_{i}.\boldsymbol{\vec{\alpha}}%
_{i}}\boldsymbol{\vec{\alpha}}_{i}.\boldsymbol{\vec{\alpha}}_{j}$ the Cartan
matric of $\mathcal{G}_{c}$ and a scalar potential as follows 
\begin{equation}
\mathcal{V}_{Toda}\left( \phi _{1},...,\phi _{r}\right)
=\sum_{i=1}^{r}\kappa _{i}\exp \left( \sum_{j=1}^{r}C_{ij}\phi _{j}\right)
\end{equation}%
For the case of sl$\left( r+1\right) $, the $C_{ij}$ is a symmetric $r\times
r$ matrix reading as.%
\begin{equation}
C_{ij}=\left( 
\begin{array}{cccc}
2 & -1 & 0 & 0 \\ 
-1 & 2 & \ddots & 0 \\ 
0 & \ddots & \ddots & -1 \\ 
0 & 0 & -1 & 2%
\end{array}%
\right) _{r\times r}=\boldsymbol{\vec{\alpha}}_{i}.\boldsymbol{\vec{\alpha}}%
_{j}  \label{td}
\end{equation}%
\textrm{For more details; see appendix section}. Second, the scalar
potential $\mathcal{V}\left( \phi \right) $ is highly non linear (non
polynomial) and its minimum takes place at $\phi \rightarrow -\infty $
making the quantisation of the 2d Liouville field $\phi $ difficult to do by
using the standard canonical QFT manner \textrm{\cite{1KC,2KC,3KC}}. \newline
A similar equation to (\ref{s0}) and related expressions can be also written
down for the 2d euclidian plane $\mathbb{R}^{2}$ parameterised $X^{m}=\left(
x,y\right) $; and the same comments given above are still valid. By using $%
\mathbb{R}^{2}\sim \mathbb{C}$, we can also use the complex coordinate $\xi
=x+iy$ and its conjugate $\bar{\xi}=x-iy$ to deal with%
\begin{equation}
\frac{\partial ^{2}\phi }{\partial \xi \partial \bar{\xi}}+\kappa e^{2\phi
}=0  \label{1f}
\end{equation}%
Here, the real Liouville field $\phi =\phi \left( \xi ,\bar{\xi}\right) $ is
a function of the complex variable $\xi $ and its conjugate that may be
formally denoted like $\xi \equiv \xi ^{+}$ and $\bar{\xi}\equiv \xi ^{-}$
where now $\pm $ stand for U$\left( 1\right) \simeq SO\left( 2\right) $
charges. As the two local coordinates $\mathrm{\rho }^{\pm }$ and $\xi ^{\pm
}$ may be related by a Wick rotation of the time direction (say $y=i\sigma $%
), the general analysis of the two flat 2d geometries is quite similar; so
one may treat both of the Lorentzian and euclidian equations collectively
like 
\begin{equation}
\frac{\partial ^{2}\phi }{\partial X^{+}\partial X^{-}}+\kappa e^{2\phi }=0
\label{s1}
\end{equation}%
where $X^{\pm }$ designate either $\mathrm{\rho }^{\pm }$ or $\xi ^{\pm }$
and parameterise a real surface $\Sigma $ as in eq(\ref{s2}). In what
follows, we shall focus on the euclidian 2d geometry with $X^{\pm }=x\pm iy$%
, we sometimes use also $\left( \xi ,\bar{\xi}\right) $ to designate $\left(
X^{+},X^{-}\right) $ in order to simplify the notations. To fix ideas,
notice also that the 2d space $\Sigma $ may viewed as the analogous one in
the 4d space $\mathbb{M}_{4}=\Sigma \times \mathcal{C}$ used in the CWY
theory.

\subsection{Special properties of eq(\protect \ref{s1})}

First, recall that the properties of 2d Liouville theory are very well known
and have been extensively studied in the mathematical physics literature
from several views \textrm{\cite{1L}-\cite{8L}}; so we will target in what
follows only on those useful aspects directly relevant for our present
construction; three of these aspects concern particularly: $\left( i\right) $
the infinite dimensional conformal symmetry of eq(\ref{s1}), $\left(
ii\right) $ the factorisation of the modulus $\kappa $ of the Liouville
theory as the product of two terms; and $\left( iii\right) $ the classical
solvability of (\ref{s0}-\ref{s1}).

\  \  \  \  \  \newline
$\mathbf{1)}$ \emph{Conformal symmetry of} (\ref{s1}) \newline
Under the holomorphic coordinate change $\xi \rightarrow \xi ^{\prime
}=f\left( \xi \right) $, the Liouville eq(\ref{s1}) remains invariant
provided the scalar potential $e^{2\phi }$ transforms in same manner like $%
\frac{\partial ^{2}\phi }{\partial \xi \partial \bar{\xi}}$, that is 
\begin{equation}
\frac{\partial ^{2}\phi ^{\prime }}{\partial \xi ^{\prime }\partial \bar{\xi}%
^{\prime }}=\left \vert \frac{\partial f}{\partial \xi }\right \vert ^{-2}%
\frac{\partial ^{2}\phi }{\partial \xi \partial \bar{\xi}}  \label{c1}
\end{equation}%
and%
\begin{equation}
e^{2\phi ^{\prime }}=\left \vert \frac{\partial f}{\partial \xi }\right
\vert ^{-2}e^{2\phi }  \label{c2}
\end{equation}%
Holomorphy of $f\left( \xi \right) $ and the symmetry of eq(\ref{s1}) lead
therefore to the following relationship between $\phi $ and $\phi ^{\prime
}, $ 
\begin{equation}
2\phi ^{\prime }\left( \xi ^{\prime },\bar{\xi}^{\prime }\right) =2\phi
\left( \xi ,\bar{\xi}\right) -\ln \left \vert \frac{\partial f}{\partial \xi 
}\right \vert ^{2}  \label{ct}
\end{equation}%
defining the conformal transformation of the Liouville field $\phi $. Quite
similar relations can be also written down for 2d Toda fields.

\  \  \  \  \  \newline
$\mathbf{2)}$ \emph{Factoring the coupling} $\kappa $ \emph{in eq} (\ref{s1}%
) \newline
From now on, we will think on the real parameter\textrm{\ }$\kappa $ of the
Liouville theory as given by the product of two non zero real numbers $%
\alpha $ and $\beta $ as follows 
\begin{equation}
\kappa =\mathrm{\alpha }\times \mathrm{\beta }>0  \label{ka}
\end{equation}%
This factorisation is important in figuring out novel properties on the
integrability of the equation as well as its embedding into the CWY theory.
For example, the two $\alpha $ and $\beta $ parameters will be used in
building a general form of the Lax pair $L_{m}=\left( L_{+},L_{-}\right) $
underlying the linearization of (\ref{s1}). In terms of this pair of field
operators, the Liouville equation can be brought to the form \cite{1M}-\cite%
{2N},%
\begin{equation}
\partial _{+}L_{-}-\partial _{-}L_{+}+\left[ L_{+},L_{-}\right] =0
\label{lv}
\end{equation}%
where the relations between the Lax pair $\left( L_{+},L_{-}\right) $ and
the Liouville field $\phi $ will be given later on.

\  \  \  \  \newline
$\mathbf{3)}$ \emph{Classical integrability}\newline
Before formulating the solvability of the Liouville equation as in eq(\ref%
{lv}), recall that a particular solution of the Liouville equation (\ref{s1}%
) can be explicitly written down. Up to the conformal transformation (\ref%
{ct}) that leaves the Liouville field action invariant, it is not difficult
to check that 
\begin{equation}
\phi =\ln \left( \frac{1}{1+\kappa \xi \bar{\xi}}\right) \qquad ,\qquad
e^{2\phi }=\frac{1}{\left( 1+\kappa \xi \bar{\xi}\right) ^{2}}  \label{fd}
\end{equation}%
is an exact solution of (\ref{s1}). From this expression, we have 
\begin{equation}
\partial _{\xi }\phi =-\frac{\kappa \bar{\xi}}{1+\kappa \xi \bar{\xi}}\qquad
,\qquad \partial _{\bar{\xi}}\partial _{\xi }\phi =-\kappa \left( 1+\kappa
\xi \bar{\xi}\right) ^{-2}
\end{equation}%
Notice that for the limit of small $\kappa \xi \bar{\xi}$, say near the
origin $\xi \rightarrow 0$, the Liouville field $\phi \sim -\kappa \xi \bar{%
\xi}\rightarrow 0$; so the linear term $\frac{\partial ^{2}\phi }{\partial
\xi \partial \bar{\xi}}$ behaves as $-\kappa \left( 1-2\kappa \xi \bar{\xi}%
\right) \rightarrow -\kappa $ in the same manner as the opposite of $\kappa
e^{2\phi }$ which behaves like $\kappa \left( 1-2\kappa \xi \bar{\xi}\right)
\rightarrow \kappa $. For the large limit $\left \vert \kappa \xi \bar{\xi}%
\right \vert >>1,$ say near $\left \vert \xi \right \vert \rightarrow \infty 
$, the Liouville field $\phi \sim \ln \frac{1}{\kappa \xi \bar{\xi}}$ goes
to $-\infty $. In this case, the linear term $\frac{\partial ^{2}\phi }{%
\partial \xi \partial \bar{\xi}}$ behaves as $-\frac{\kappa }{\left( \kappa
\xi \bar{\xi}\right) ^{2}}$ and goes to $0$ in the same way as $\kappa
e^{2\phi }$ which behaves as $\frac{\kappa }{\left( \kappa \xi \bar{\xi}%
\right) ^{2}}$ and then tends to zero as well. A quite general form of the
solution of the Liouville equation is given by $\phi =\frac{1}{2}\ln \left( 
\frac{\left \vert \partial _{\xi }f\right \vert ^{2}}{\left( 1+\kappa
\left
\vert f\right \vert ^{2}\right) ^{2}}\right) $ with $f=f\left( \xi
\right) $.

\section{More on Lax formulation}

Here, we study some useful properties of the Lax pair $\left(
L_{+},L_{-}\right) $ appearing in eq(\ref{lv}). First, we describe the link
between the Lax equation (\ref{lv}) and the gauge field eqs(\ref{29}) in the
CWY theory. Then, we derive the relation between the 2d Liouville field $%
\phi \left( X^{\pm }\right) $ and the $L_{\pm }$ Lax pair. We also study the
set $\boldsymbol{H}\times Hol\left( \Sigma \right) $ of local symmetries of
the Lax pair; this set is contained into $SL\left( 2\right) \times
Diff\left( \Sigma \right) $ and is obtained by solving constraint relations
on some components of the CWY gauge connection imposed by the embedding of
Liouville equation. The construction given here below for Liouville model
applies as well to the full set of the finite 2d Toda QFT$_{2}$'s.

\subsection{Lax pair as a particular CWY gauge configuration}

The Lax pair $\left( L_{+},L_{-}\right) $, which satisfy the Lax equation eq(%
\ref{lv}) linearising the Liouville equation, is very suggestive. Comparing (%
\ref{lv}) with the first relation of eqs(\ref{29}) in the CWY theory namely%
\begin{equation}
\mathcal{F}_{\left[ +-\right] }=\partial _{+}\mathcal{A}_{-}-\partial _{-}%
\mathcal{A}_{+}+\left[ \mathcal{A}_{+},\mathcal{A}_{-}\right] =0  \label{14}
\end{equation}%
one may think of eq(\ref{s1}) and then of eq(\ref{lv}) as following from the
4d eq(\ref{14}) by imposing constraints on some components of $\mathcal{A}%
_{\pm }$ along the sl$\left( 2\right) $ fiber directions. In this view, the $%
L_{\pm }$ operators can be imagined as describing a particular non abelian
gauge configuration solving the field equation of the CWY gauge field namely 
\begin{equation}
\mathcal{F}_{\mu \nu }=\mathcal{F}_{\mu \nu }\left( X^{\pm },\zeta ,\bar{%
\zeta}\right) =0
\end{equation}%
This antisymmetric $\mathcal{F}_{\mu \nu }$ tensor splits in the $X^{\pm }$
and $\bar{\zeta}$ directions as follows%
\begin{equation}
\mathcal{F}_{\mu \nu }=\left( 
\begin{array}{cc}
\mathcal{F}_{mn} & \mathcal{F}_{m\bar{\zeta}} \\ 
\mathcal{F}_{\bar{\zeta}n} & 0%
\end{array}%
\right)
\end{equation}%
with $\mathcal{F}_{mn}=\varepsilon _{mn}\mathcal{F}_{\left[ +-\right] }$ as
in (\ref{14}) and $\mathcal{F}_{m\bar{\zeta}}=\left( \mathcal{F}_{+\bar{\zeta%
}},\mathcal{F}_{-\bar{\zeta}}\right) $. It splits as well with respect to
the sl$\left( 2\right) $ fiber directions like $\mathcal{F}_{\mu \nu }=t_{a}%
\mathcal{F}_{\mu \nu }^{a}$. By expanding the partial 1-form gauge
connection $\mathcal{A}=\mathcal{A}\left( X^{\pm },\zeta ,\bar{\zeta}\right) 
$ along the $dX^{\pm }$ and $d\bar{\zeta}$ dimensions like 
\begin{equation}
\mathcal{A}=\mathcal{A}_{+}dX^{+}+\mathcal{A}_{-}dX^{-}+\mathcal{A}_{\bar{%
\zeta}}d\bar{\zeta}
\end{equation}%
and setting the differential 
\begin{equation}
d\bar{\zeta}=0
\end{equation}%
by demanding to the variable $\bar{\zeta}$ to sit at some fixed constant
value --- for example by setting $\bar{\zeta}=0$ ---, the above expansion
reduces to 2d space gauge connection $A=A_{+}dX^{+}+A_{-}dX^{-}$ with $%
A_{\pm }=A_{\pm }\left( X^{\pm }\right) $. So, one can imagine the Lax pair $%
L_{\pm }=L_{\pm }\left( X^{\pm }\right) $ of the 2d Liouville theory as
contained in $A_{\pm }\left( X^{\pm }\right) $ as follows 
\begin{equation}
\begin{tabular}{lll}
$L_{+}$ & $\subset $ & $A_{+}\left( X^{\pm }\right) $ \\ 
$L_{-}$ & $\subset $ & $A_{-}\left( X^{\pm }\right) $%
\end{tabular}
\label{in}
\end{equation}%
Notice that the $A_{\pm }$ are reductions of the CWY vector potential
components $\mathcal{A}_{\pm }=\mathcal{A}_{\pm }\left( X^{\pm },\zeta ,\bar{%
\zeta}\right) $ down to 2d as they are given by 
\begin{equation}
\begin{tabular}{lll}
$A_{+}\left( X^{\pm }\right) $ & $=$ & $\left. \mathcal{A}_{+}\left( X^{\pm
},\zeta ,\bar{\zeta}\right) \right \vert _{\zeta =\bar{\zeta}=0}$ \\ 
$A_{-}\left( X^{\pm }\right) $ & $=$ & $\left. \mathcal{A}_{-}\left( X^{\pm
},\zeta ,\bar{\zeta}\right) \right \vert _{\zeta =\bar{\zeta}=0}$%
\end{tabular}%
\end{equation}%
So they are zero modes on the complex curve $\mathcal{C}$ of the 4d space $%
\mathbb{M}_{4}=\Sigma \times \mathcal{C}$, and they satisfy the vanishing 2d
space curvature condition%
\begin{equation}
\partial _{+}A_{-}-\partial _{-}A_{+}+\left[ A_{+},A_{-}\right] =0
\label{cv}
\end{equation}%
that follows from eq(\ref{14}) by fixing $\bar{\zeta}$ to a constant. The
inclusion $\subset $ symbol in eqs(\ref{in}) means that $L_{\pm }$ are given
by pieces of the Lie algebra expansion of the non abelian vector potential $%
A_{\pm }=\sum_{a}t_{a}A_{\pm }^{a}$. The $t_{a}$'s are the generators of the 
$\mathcal{G}_{c}$ Lie algebra satisfying the commutation relation%
\begin{equation}
\left[ t_{a},t_{b}\right] =f_{abc}t_{c}
\end{equation}%
Put differently, the 2- dimensional $L_{\pm }$ Lax operators of Liouville
equation can be recovered from the $\mathcal{A}_{\pm }$ components of CWY
gauge connection by taking $\mathcal{G}_{c}=$ $sl\left( 2\right) $ and
imposing constraints on some of the direction of propagation of the gauge
potential $\mathcal{A}_{\mu }$ in gauge fiber bundle. These constraints will
be derived in what follows; but after deriving the explicit expression of $%
L_{\pm }$ in terms of the Liouville field.

\subsection{From Liouville field to Lax pair $L_{\pm }$}

First, we construct the relationship between the Liouville field $\phi $ and
the Lax pair $\left( L_{+},L_{-}\right) $ by using two manners (top- down
and bottom- up) to get more insight into the construction: $\left( i\right) $
top- down: this is a short and somehow heuristic manner using specific
features to build easily $L_{\pm }\left( \phi \right) $; it applies to 2d
models like the finite Toda QFT$_{2}$s considered in this study. $\left(
ii\right) $ bottom- up: this is a systematic manner based on general
arguments and which may be used for generic cases. Then, we give some
properties on the link $L_{\pm }=L_{\pm }\left( \phi \right) $ as well as
the interpretation of $L_{+},L_{-}$ as particular components of a non
abelian gauge potential $A_{\pm }=t_{a}A_{\pm }^{a}$.

\subsubsection{Relationship between $\protect \phi $ and $L_{\pm }$}

To begin, notice that the link between the Liouville field $\phi $ and the
Lax pair defines the transformations $L_{\pm }=L_{\pm }\left( \phi \right) $
one has to do in order to linearise the Liouville equation. The use of a
pair $L_{+}$ and $L_{-}$ of variables at the place of the unique field
variable $\phi $ is the price to pay for linearization. The relationship
between $\phi $ and $L_{\pm }$ have the following dependence 
\begin{equation}
L_{+}=L_{+}\left( \phi ,\partial \phi ;t_{a}\right) \qquad ,\qquad
L_{-}=L_{-}\left( \phi ,\partial \phi ;t_{a}\right)  \label{exc}
\end{equation}%
where $\partial \phi $ stands for the 2d gradient of the Liouville field and 
$t_{a}$ for the sl$\left( 2\right) $ generators to be taken in the Cartan
basis; i.e: $t_{a}\equiv \left( h,E^{-},E^{+}\right) $.

\  \  \ 

\emph{1) Heuristic derivation of L}$_{\pm }$\newline
By thinking of the Liouville equation eq(\ref{s1}) as a matrix relation
valued in the diagonal h- direction of $sl\left( 2\right) $ like 
\begin{equation}
\left( \frac{\partial ^{2}\phi }{\partial X^{+}\partial X^{-}}+\kappa
e^{2\phi }\right) h=0  \label{xx}
\end{equation}%
with coupling $\kappa =\mathrm{\alpha \beta }$, and equating this matrix
with the Lax equation $\partial _{+}L_{-}-\partial _{-}L_{+}+\left[
L_{+},L_{-}\right] =0$ of eq(\ref{lv}), we can derive the explicit
expressions of the $L_{\pm }$ Lax pair in terms of the constants $\mathrm{%
\alpha }$, $\mathrm{\beta }$; the Liouville field $\phi =\phi \left(
X^{+},X^{-}\right) $ and its gradient $\partial _{\pm }\phi $ with $\partial
_{\pm }=\frac{1}{2}(\frac{\partial }{\partial x}\mp i\frac{\partial }{%
\partial y})$. Using the following commutation relations 
\begin{equation}
\left[ h,E^{\pm }\right] =\pm 2E^{\pm }\qquad ,\qquad \left[ E^{+},E^{-}%
\right] =h
\end{equation}%
and thinking of the non linear term $\kappa e^{2\phi }h$ in the Liouville
field as intimately related with the commutator $\left[ L_{+},L_{-}\right] $
especially with $\left[ \mathrm{\alpha }E^{+},\mathrm{\beta }e^{2\phi }E^{-}%
\right] $; i.e:%
\begin{equation}
\kappa e^{2\phi }h=\left[ \mathrm{\alpha }E^{+},\mathrm{\beta }e^{2\phi
}E^{-}\right]
\end{equation}%
one can easily check that the following expressions give a realisation of
the Lax operators in terms of $\phi $ and $h,E^{+},E^{-},$%
\begin{eqnarray}
L_{+} &=&\left( \partial _{+}\phi \right) h-\alpha E^{+}  \notag \\
L_{-} &=&\beta e^{2\phi }E^{-}  \label{4}
\end{eqnarray}%
Putting these quantities back into $L_{x}=L_{+}+L_{-}$ and $L_{y}=\frac{1}{i}%
\left( L_{+}-L_{-}\right) $, we obtain the following complex quantities 
\begin{eqnarray}
L_{x} &=&\left( \frac{\partial \phi }{\partial x}-i\varepsilon _{xy}\frac{%
\partial \phi }{\partial y}\right) \frac{h}{2}-\alpha E^{+}+\beta e^{2\phi
}E^{-}  \notag \\
L_{y} &=&\left( \frac{\partial \phi }{\partial y}+i\varepsilon _{yx}\frac{%
\partial \phi }{\partial x}\right) h+i\alpha E^{+}+i\beta e^{2\phi }E^{-}
\label{xy}
\end{eqnarray}%
where we have used $\varepsilon _{xy}=-\varepsilon _{yx}=1$ and $\varepsilon
_{xx}=\varepsilon _{yy}=0$. Substituting eqs(\ref{4}) back into the Lax
equation, we rediscover (\ref{xx}).

\  \  \ 

\emph{2) Rigourous derivation of L}$_{\pm }$\newline
A quite rigourous manner to get the expressions in (\ref{4}) compared to the
previous heuristic one is to proceed as follows: $\left( i\right) $ start
from eq(\ref{cv}) describing the condition of vanishing curvature $F_{\left[
+-\right] }=\varepsilon _{+-}\boldsymbol{F}=0$ of a generic non abelian sl$%
\left( 2\right) $ vector potential $A_{m}=\left( A_{+},A_{-}\right) ,$ 
\begin{equation}
\boldsymbol{F}=\partial _{+}A_{-}-\partial _{-}A_{+}+\left[ A_{+},A_{-}%
\right]  \label{1a}
\end{equation}%
and $\left( ii\right) $ look for a particular solution that fit with the
Liouville equation. In this manner of doing one has to impose constraints on
some components of the vector potential $A_{m}^{a}$; this may be achieved in
two steps as follows:

\begin{itemize}
\item \emph{step 1}: project the vanishing curvature matrix condition $%
\boldsymbol{F}=0$ along the three directions of $sl\left( 2\right) $ like $%
Tr\left( t^{a}\boldsymbol{F}\right) =\boldsymbol{F}^{a}=0$ with 
\begin{equation}
\boldsymbol{F}^{a}=\partial _{+}A_{-}^{a}-\partial
_{-}A_{+}^{a}+f^{abc}A_{+}^{b}A_{-}^{c}
\end{equation}%
The resulting three scalar conditions are nicely formulated by using the
Cartan basis of $sl\left( 2\right) $ as follows:%
\begin{eqnarray}
Tr\left( \frac{h}{2}\boldsymbol{F}\right) &=&\boldsymbol{F}^{0}=0  \notag \\
Tr\left( E^{+}\boldsymbol{F}\right) &=&\boldsymbol{F}^{+}=0  \label{3e} \\
Tr\left( E^{-}\boldsymbol{F}\right) &=&\boldsymbol{F}^{-}=0  \notag
\end{eqnarray}%
with neutral $\boldsymbol{F}^{0}$ component given by%
\begin{equation}
\boldsymbol{F}^{0}=\partial _{+}A_{-}^{0}-\partial
_{-}A_{+}^{0}+A_{+}^{-}A_{-}^{+}-A_{+}^{+}A_{-}^{-}  \label{01}
\end{equation}%
and two charged $\boldsymbol{F}^{\pm }$ ones like%
\begin{eqnarray}
\boldsymbol{F}^{+} &=&\partial _{+}A_{-}^{+}-\partial _{-}A_{+}^{+}+2\left(
A_{+}^{+}A_{-}^{0}-A_{+}^{0}A_{-}^{+}\right)  \notag \\
\boldsymbol{F}^{-} &=&\partial _{+}A_{-}^{-}-\partial _{-}A_{+}^{-}+2\left(
A_{+}^{0}A_{-}^{-}-A_{+}^{-}A_{-}^{0}\right)  \label{2}
\end{eqnarray}

\item \emph{step 2}: solve the two charged equations $\boldsymbol{F}^{\pm
}=0 $ in terms of a real scalar field $\phi $; and put the obtained solution
back into (\ref{01}). However, the solving of (\ref{2}) should be such that
one ends with the Liouville equation; this requires imposing appropriate
constraints on some of the components appearing the following expansion 
\begin{equation}
A_{m}=A_{m}^{0}h+A_{m}^{-}E^{+}+A_{m}^{+}E^{-}  \label{p}
\end{equation}%
The determination of the appropriate constraints on the $A_{m}^{0,\pm }$'s
can be motivated from the structure of the Liouville equation which
indicates that we should have an $\boldsymbol{F}^{0}$ equation containing
two terms like for instance%
\begin{equation}
\boldsymbol{F}^{0}=\partial _{-}A_{+}^{0}-A_{+}^{-}A_{-}^{+}=0  \label{fr}
\end{equation}%
The first $\partial _{-}A_{+}^{0}$ term in above $\boldsymbol{F}^{0}$ is
needed to generate the laplacian $\partial _{-}\partial _{+}\phi $; and the
second one namely $A_{+}^{-}A_{-}^{+}$ is needed to recover the contribution
coming from the scalar potential $\kappa e^{2\phi }$. By comparing (\ref{fr}%
) with (\ref{01}), we end with the following constraint relations we have to
impose 
\begin{equation}
\begin{tabular}{lll}
$A_{+}^{+}$ & $=$ & $0$ \\ 
$A_{-}^{0}$ & $=$ & $0$ \\ 
$A_{-}^{-}$ & $=$ & $0$%
\end{tabular}
\label{24}
\end{equation}%
Putting these constraints back into (\ref{2}), we end with reduced charged $%
\boldsymbol{F}^{\pm }$ curvatures that we have to solve in terms of the
Liouville field $\phi $ and other parameters,%
\begin{eqnarray}
\boldsymbol{F}^{-} &=&\frac{\partial A_{+}^{-}}{\partial X^{-}}=0  \notag \\
\boldsymbol{F}^{+} &=&\left( \frac{\partial }{\partial X^{+}}%
-2A_{+}^{0}\right) A_{-}^{+}=0  \label{rf}
\end{eqnarray}%
The first relation of (\ref{rf}) namely $\frac{\partial A_{+}^{-}}{\partial
X^{-}}=0$ is solved by $A_{+}^{-}=A_{+}^{-}\left( X^{+}\right) $; but
because of conformal symmetry (\ref{ct}), it can be thought of as just given
by the constant $A_{+}^{-}=-\alpha $. The second relation can be also solved
exactly as follows 
\begin{equation}
A_{+}^{0}=\frac{\partial \phi }{\partial X^{+}}\qquad ,\qquad
A_{-}^{+}=\beta e^{2\phi }  \label{1b}
\end{equation}%
By putting these solutions back into (\ref{fr}), we get $\frac{\partial
^{2}\phi }{\partial X^{-}\partial X^{+}}+\alpha \beta e^{2\phi }=0$
coinciding exactly with the Liouville equation.
\end{itemize}

\subsubsection{More on the link $L_{\pm }/A_{\pm }$}

The realisation of the Lax operators $L_{\pm }$ in terms of $\phi $ and the
sl$\left( 2\right) $ generators given by eqs(\ref{4}) has some particular
features that are interesting for the study of the embedding of the
Liouville equation into the CWY theory with SL$\left( 2\right) $ gauge
symmetry. Here, we want to describe five of these remarkable properties;
they are as listed here below.

\begin{description}
\item[ $i)$] \emph{complexified Lie algebra }$sl\left( 2\right) $:\newline
The Lax operators $L_{+}$ and $L_{-}$ linearising Liouville equation are non
hermitian operators since from (\ref{4}) we learn 
\begin{equation}
\left( L_{+}\right) ^{\dagger }\neq L_{-}\qquad ,\qquad \left( L_{-}\right)
^{\dagger }\neq L_{+}
\end{equation}%
Similarly, we have from (\ref{xy}), 
\begin{equation}
\left( L_{x}\right) ^{\dagger }\neq L_{x}\qquad ,\qquad \left( L_{y}\right)
^{\dagger }\neq L_{y}
\end{equation}%
The non hermiticity property of the Lax formalism may be also exhibited by
using the $2\times 2$ matrix representation of the generators $\left(
h,E^{\pm }\right) ;$ we have%
\begin{equation}
L_{+}=\left( 
\begin{array}{cc}
\partial _{+}\phi & -\alpha \\ 
0 & -\partial _{+}\phi%
\end{array}%
\right) \qquad ,\qquad L_{-}=\left( 
\begin{array}{cc}
0 & 0 \\ 
\beta e^{2\phi } & 0%
\end{array}%
\right)
\end{equation}%
The complex behavior of the $L_{x}$ and $L_{y}$ operators fit well with the
formal action of the CWY theory requiring a complexified gauge symmetry $%
\mathcal{G}_{c}$ which in the Liouville model is given by $sl\left( 2\right) 
$.

\item[ $ii)$] \emph{constrained vector potential }$B_{m}$\emph{:}\newline
By comparing eq(\ref{4}) with the expansion of a generic vector potential $%
A_{\pm }$ along the $sl\left( 2\right) $ directions namely 
\begin{equation}
\begin{tabular}{lll}
$A_{+}$ & $=$ & $A_{+}^{0}h+A_{+}^{-}E^{+}+A_{+}^{+}E^{-}$ \\ 
$A_{-}$ & $=$ & $A_{-}^{0}h+A_{-}^{-}E^{+}+A_{-}^{+}E^{-}$%
\end{tabular}%
\end{equation}%
we end with the relationships%
\begin{equation}
\begin{tabular}{lll}
$A_{+}^{0}$ & $=$ & $\partial _{+}\phi $ \\ 
$A_{+}^{-}$ & $=$ & $-\alpha $ \\ 
$A_{-}^{+}$ & $=$ & $\beta e^{2\phi }$%
\end{tabular}
\label{23}
\end{equation}%
together with the following constraints%
\begin{equation}
A_{+}^{+}=0\qquad ,\qquad A_{-}^{0}=0\qquad ,\qquad A_{-}^{-}=0
\end{equation}%
which are precisely the ones obtained by the rigourous manner in deriving $%
L_{\pm }=L_{\pm }\left( \phi \right) $. These constraint relations play an
important role in our construction; they show that the $A_{+}^{+}$ component
of the left chirality $A_{+}$ of vector potential $A_{\pm }$ should not
propagate in the $E^{-}$ direction of the $sl\left( 2\right) $ gauge bundle;
and similarly the $A_{-}^{0},A_{-}^{-}$ $A_{-}$ components of the right
chirality $A_{-}$ which should not spread in the $h$ and $E^{+}$ directions;
this feature is illustrated on the figure \ref{fa}. 
\begin{figure}[tbph]
\begin{center}
\includegraphics[scale=0.63]{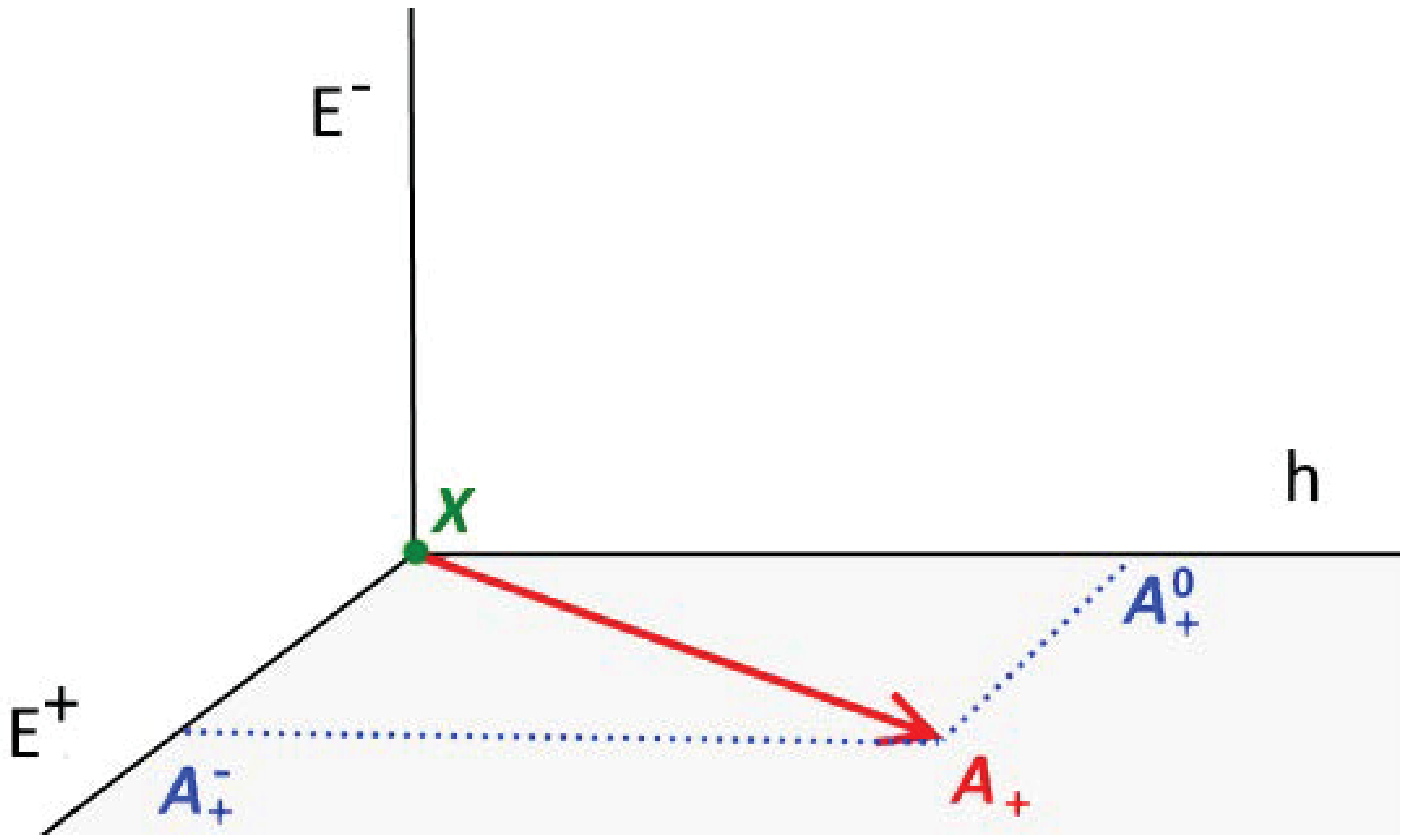}\quad %
\includegraphics[scale=0.63]{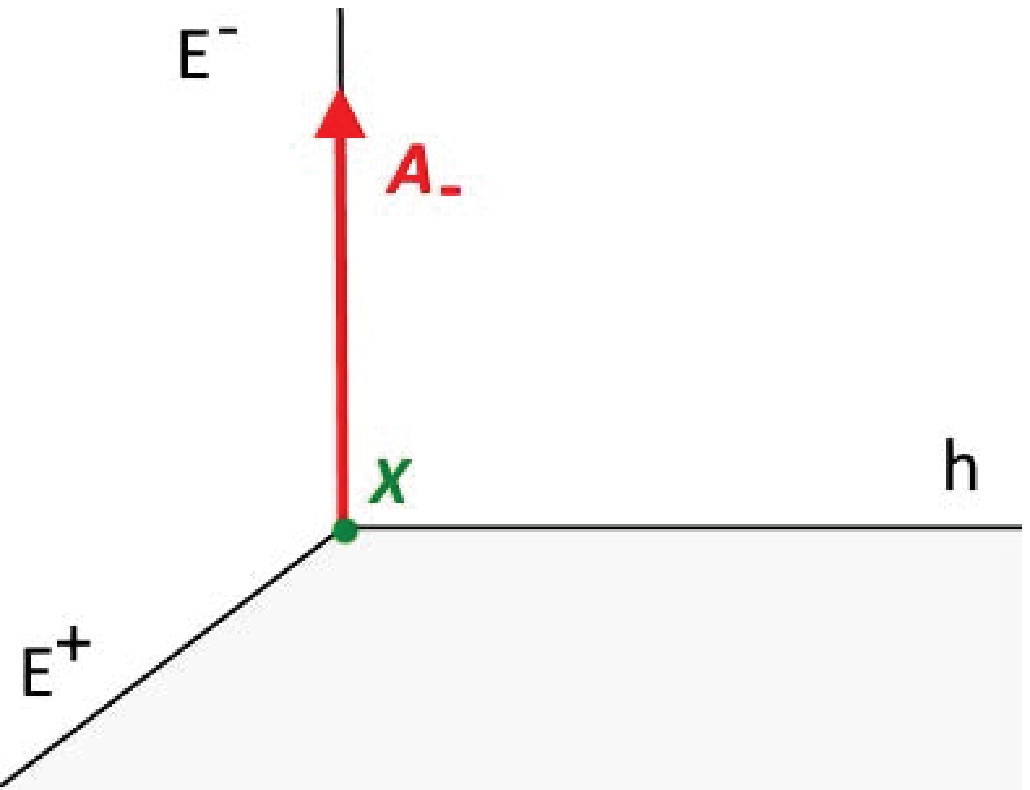}
\end{center}
\par
\vspace{0cm}
\caption{A schematic representation of the propagation directions in the SL$%
\left( 2\right) $ gauge bundle of the non abelian vector potential $A_{\pm
}=A_{\pm }\left( X\right) $ inducing the Liouville equation. The direction
of $A_{+}=A_{+}\left( X\right) $ is shown on the left panel; it has no
component on $E^{-}$. The direction of $A_{-}=A_{-}\left( X\right) $ is
shown on the right panel; it has one non zero component along the $E^{-}$
direction. The $\left( h,E^{+},E^{-}\right) $ is the basis of $sl\left(
2\right) $ at each $X=\left( X^{+},X^{-}\right) $; and the projection of $%
A_{m}=A_{m}\left( X\right) $ along a given direction axis $\boldsymbol{t}$
is given by $Tr\left[ \boldsymbol{t}A_{m}\right] $.}
\label{fa}
\end{figure}
This means that the gauge potential associated with the Liouville theory is
not a generic vector potential; but a constrained $B_{m}$ vector potential
given by the following restriction 
\begin{equation}
B_{m}=\eta _{m-}\left( A_{+}^{0}h+A_{+}^{-}E^{+}\right) +\eta
_{m+}A_{-}^{+}E^{-}  \label{bm}
\end{equation}%
In section 5, we will give more details on the properties of this non
abelian vector potential; see for instance eq(\ref{ba}) and eq(\ref{LC}).

\item[$iii)$] \emph{behavior of }$L_{\pm }$\emph{\ in the limit }$\kappa
X^{+}X\rightarrow \infty $:\newline
By using eq(\ref{fd}) giving an exact solution of the Liouville equation, up
to the conformal transformation (\ref{ct}) namely $\tilde{\phi}=\phi -\frac{1%
}{2}\ln \left \vert \frac{\partial \tilde{X}^{+}}{\partial X^{+}}%
\right
\vert ^{2}$ with $\tilde{X}^{+}=\tilde{X}^{+}\left( X^{+}\right) $,
we can express the Lax pair as%
\begin{eqnarray}
L_{+} &=&\frac{\kappa X^{-}}{1+\kappa X^{+}X^{-}}h-\alpha E^{+}  \notag \\
L_{-} &=&\frac{\beta }{\left( 1+\kappa X^{+}X^{-}\right) ^{2}}E^{-}
\end{eqnarray}%
and then the corresponding non zero components of the non abelian vector
potential like 
\begin{eqnarray}
A_{+}^{0} &=&-\frac{\kappa X^{-}}{1+\kappa X^{+}X^{-}}  \notag \\
A_{+}^{-} &=&-\frac{\kappa }{\beta } \\
A_{-}^{+} &=&\frac{\beta }{\left( 1+\kappa X^{+}X^{-}\right) ^{2}}  \notag
\end{eqnarray}%
These relations teach us that in large limit 
\begin{equation}
\kappa X^{+}X^{-}=\kappa \left( x^{2}+y^{2}\right) >>1
\end{equation}%
the non abelian vector potential $%
A_{m}=A_{m}^{0}h+A_{m}^{-}E^{+}+A_{m}^{+}E^{-}$ in Liouville theory tends
towards the constant $-\frac{\kappa }{\beta }E^{+}\eta _{m-}$ with metric
components $\eta _{+-}=\eta _{-+}=1$ and $\eta _{++}=\eta _{--}=0$.

\item[ $iv)$] \emph{reduced gauge symmetry }$\boldsymbol{H}\subset SL\left(
2\right) $:\newline
Because of the constraints (\ref{24}) on components of the non abelian gauge
connection, the usual SL$\left( 2\right) $ gauge transformation of the
vector potential namely 
\begin{equation}
A_{m}\qquad \rightarrow \qquad A_{m}^{\prime }=g^{-1}A_{m}g+g^{-1}\partial
_{m}g
\end{equation}%
gets reduced down to a subgroup of SL$\left( 2\right) $\ that preserve (\ref%
{24}). Indeed for the constraint eqs(\ref{24}) to be invariant, the gauge
transformation 
\begin{equation}
g=e^{\vartheta }\qquad ,\qquad g=g\left( X^{+},X^{-}\right) \qquad ,\qquad
\vartheta =\vartheta \left( X^{+},X^{-}\right)
\end{equation}%
with $\vartheta =\vartheta ^{0}h+\vartheta ^{-}E^{+}+\vartheta ^{+}E^{-}$
should be in a subset $\boldsymbol{H}$ of $SL\left( 2\right) $\ whose
elements satisfy the following conditions%
\begin{eqnarray}
Tr\left( E^{+}g^{-1}\partial _{+}g\right) &=&0  \notag \\
Tr\left( E^{-}g^{-1}\partial _{-}g\right) &=&0  \label{eeh} \\
Tr\left( hg^{-1}\partial _{-}g\right) &=&0  \notag
\end{eqnarray}%
Clearly, this subset $\boldsymbol{H}$ contains the abelian subgroup of $%
SL\left( 2\right) $\ whose elements $g=\exp \left( \vartheta ^{0}h\right) $
with local parameter $\vartheta ^{0}$ a holomorphic function in the complex
variable $X^{+}$,%
\begin{equation}
\frac{\partial \vartheta ^{0}}{\partial X^{-}}=0\qquad \Rightarrow \qquad
\vartheta ^{0}=\vartheta ^{0}\left( X^{+}\right)
\end{equation}%
This is because the two first relations in (\textrm{\ref{eeh}}) are
trivially solved due to $Tr\left( E^{+}h\right) =0=Tr\left( E^{-}h\right) $;
while the third one requires $\frac{\partial \vartheta ^{0}}{\partial X^{-}}%
=0$ since $Tr\left( hh\right) \neq 0$. \newline
Notice also that as far as chiral transformations are concerned, we
distinguish two subgroups $\boldsymbol{H}_{\left( +\right) }$ and $%
\boldsymbol{H}_{\left( -\right) }$\textrm{\ depending on whether }$\partial
_{-}g=0$\textrm{\ or }$\partial _{+}g=0$.\newline
\  \  \  \  \  \newline
$\ast $ \emph{Subgroup }$\boldsymbol{H}_{\left( +\right) }$\emph{:} It
corresponds to the case where $g=g\left( X^{+}\right) $ satisfying $\frac{%
\partial g}{\partial X^{-}}=0$\textrm{. In this situation, the eqs(\ref{eeh}%
)\ reduce to the first relation}%
\begin{equation}
Tr\left( E^{+}g^{-1}\partial _{+}g\right) =0  \label{eg}
\end{equation}%
since the two $Tr\left( E^{-}g^{-1}\partial _{-}g\right) $ and $Tr\left(
hg^{-1}\partial _{-}g\right) $ vanish identically due to $\frac{\partial
\vartheta ^{0}}{\partial X^{-}}=0$ and $\frac{\partial \vartheta ^{-}}{%
\partial X^{-}}=0$. The above relation (\ref{eg}) is solved by those non
abelian transformations 
\begin{equation}
g=\exp \left( \vartheta ^{0}h+\vartheta ^{-}E^{+}\right)  \label{ge}
\end{equation}%
with analytic parameters 
\begin{equation}
\vartheta ^{0}=\vartheta ^{0}\left( X^{+}\right) \qquad ,\qquad \vartheta
^{-}=\vartheta ^{-}\left( X^{+}\right)
\end{equation}%
The restriction to eq(\ref{ge}) with no dependence into $\vartheta ^{+}$
parameter follows from the fact that $g^{-1}\partial _{m}g$ is an element of
the Lie algebra; and the identity $Tr\left[ E^{+}\left( ah+bE^{+}\right) %
\right] =0$ which holds for arbitrary numbers $a$ and $b$ but not for $Tr%
\left[ E^{+}\left( ah+bE^{+}+cE^{-}\right) \right] $ which does not vanish
for $c\neq 0$. The $\boldsymbol{H}_{\left( +\right) }$ contains the the
diagonal $\exp \left[ h\vartheta ^{0}\left( X^{+}\right) \right] $.\newline
\  \  \  \  \  \newline
$\ast $ \emph{Subroup }$\boldsymbol{H}_{\left( -\right) }$\emph{: }It
corresponds to $g=g\left( X^{-}\right) $ satisfying $\frac{\partial g}{%
\partial X^{+}}=0$\textrm{. In this case, eq(\ref{eeh})\ reduce to its two
last relations seen that the first one vanishes identically}%
\begin{eqnarray}
Tr\left( E^{-}g^{-1}\partial _{-}g\right) &=&0 \\
Tr\left( hg^{-1}\partial _{-}g\right) &=&0  \notag
\end{eqnarray}%
Because of the properties $Tr\left( E^{-}E^{+}\right) \neq 0$ and $Tr\left(
hh\right) \neq 0$, it results that the solution of these constraints is
given by 
\begin{equation}
g\left( X^{-}\right) =\exp \left[ E^{-}\vartheta ^{+}\left( X^{-}\right) %
\right]
\end{equation}%
with $\frac{\partial \vartheta ^{+}}{\partial X^{+}}=0$. So, the $%
\boldsymbol{H}_{\left( -\right) }$ does not contain the diagonal $\exp \left[
h\vartheta ^{0}\left( X^{-}\right) \right] .$

\item[$v)$] \emph{holomorphic diffeomorphisms:}\newline
Because of the constraints (\ref{24}), the invariance under $Diff\left(
\Sigma \right) $ gets reduced to the invariance under holomorphic
transformations%
\begin{equation}
X^{+}\rightarrow \tilde{X}^{+}=f\left( X^{+}\right) \qquad ,\qquad
X^{-}\rightarrow \tilde{X}^{-}=\bar{f}\left( X^{-}\right)
\end{equation}%
This holomorphic feature can be derived by using the language of
differential forms on $\Sigma $; the gauge connection $A\left( X\right)
=A_{+}dX^{+}+A_{-}dX^{-},$ in coordinate frame $X^{\pm }$, transforms in $%
\tilde{X}^{\pm }$- coordinate frame into $\tilde{A}(\tilde{X})=\tilde{A}_{+}d%
\tilde{X}^{+}+\tilde{A}_{-}d\tilde{X}^{-}$. Solving the identity $A\left(
X\right) =\tilde{A}(\tilde{X})$ and substituting back into (\ref{24}), we
obtain 
\begin{eqnarray}
A_{-}^{0} &=&\frac{\partial \tilde{X}^{+}}{\partial X^{-}}\tilde{A}_{+}^{0}+%
\frac{\partial \tilde{X}^{-}}{\partial X^{-}}\tilde{A}_{-}^{0}=0  \notag \\
A_{-}^{-} &=&\frac{\partial \tilde{X}^{+}}{\partial X^{-}}\tilde{A}_{+}^{-}+%
\frac{\partial \tilde{X}^{-}}{\partial X^{-}}\tilde{A}_{-}^{-}=0 \\
A_{+}^{+} &=&\frac{\partial \tilde{X}^{+}}{\partial X^{+}}\tilde{A}_{+}^{+}+%
\frac{\partial \tilde{X}^{-}}{\partial X^{+}}\tilde{A}_{-}^{+}=0  \notag
\end{eqnarray}%
Demanding invariance under diffeomorphism of the constraint eqs(\ref{24});
that is $\tilde{A}_{-}^{0}=\tilde{A}_{-}^{-}=\tilde{A}_{+}^{+}=0$, we end
with%
\begin{eqnarray}
\frac{\partial \tilde{X}^{+}}{\partial X^{-}}\tilde{A}_{+}^{0} &=&0  \notag
\\
\frac{\partial \tilde{X}^{+}}{\partial X^{-}}\tilde{A}_{+}^{-} &=&0 \\
\frac{\partial \tilde{X}^{-}}{\partial X^{+}}\tilde{A}_{-}^{+} &=&0  \notag
\end{eqnarray}%
requiring the conditions 
\begin{equation}
\frac{\partial \tilde{X}^{+}}{\partial X^{-}}=0\qquad ,\qquad \frac{\partial 
\tilde{X}^{-}}{\partial X^{+}}=0
\end{equation}%
and then holomorphic coordinates transformations. This feature indicates
that the properties of the real surface $\Sigma $ have to be approached in
terms of properties of a complex curve.
\end{description}

\section{From Liouville to CWY}

In section 4, we have shown that the Lax pair $\left( L_{+},L_{-}\right) $
given by eq(\ref{4}), linearising the 2d Liouville equation, is in fact a
particular solution of the equations of motion (\ref{19}) describing the
dynamics of the CWY gauge connection in 4d; see the figure \ref{fa}. In this
section, we complete this analysis by studying in subsection 5.1 the
generalisation of these Lax operators to the 4d space $\Sigma \times 
\mathcal{C}$ as well as some of their properties. With this link between the
standard integrable 2d Lax formalism and the formal CWY gauge theory, we
dispose of a manner to construct observables in Toda QFT$_{2}$'s by using
quantum Wilson line operators and their generalisations. In subsection 5.2,
we build these quantum line operators for the example of Liouville theory
and show how they are characterised by a rank four tensor $\Gamma _{\mu
a}^{\nu b}$ where $\mu ,\nu $ are space indices and $a,b$ Lie algebra ones.

\subsection{4d extended Lax equations}

Using results from \cite{1F} on Costello-Witten-Yamazaki 4d topological
gauge theory, whose useful aspects to the present analysis were presented in
section 2, and focussing on the case of a 4d space $\mathbb{M}_{4}$ given by 
$\mathbb{R}^{2}\times \mathbb{C}$ parameterised by the local coordinates 
\begin{equation}
X^{M}=\left( X^{+},X^{-};\zeta ,\bar{\zeta}\right)
\end{equation}%
with $X^{\pm }=x\pm iy$ for $\Sigma =\mathbb{R}^{2}\sim \mathbb{C}$ and $%
\zeta =X_{3}+iX_{4}$ for the complex line $\mathcal{C}=\mathbb{C}$, we can
write down an extension of the 2d Lax equation on $\mathbb{R}^{2}$\ to the 4
space $\mathbb{R}^{2}\times \mathbb{C}\simeq \mathbb{C}\times \mathbb{C}$.
In this 4d extension, the usual 2d Lax equation 
\begin{equation}
\partial _{+}L_{-}-\partial _{-}L_{+}+\left[ L_{+},L_{-}\right] =0
\end{equation}%
expressed in terms of the 2d Lax pair $\left( L_{+},L_{-}\right) $ living on 
$\mathbb{R}^{2}$ with 
\begin{equation}
L_{+}=L_{+}\left( X^{+},X^{-};t_{a}\right) \qquad ,\qquad L_{-}=L_{-}\left(
X^{+},X^{-};t_{a}\right)
\end{equation}%
gets promoted to three 4d extended equations%
\begin{eqnarray}
\partial _{+}\mathcal{L}_{-}-\partial _{-}\mathcal{L}_{+}+\left[ \mathcal{L}%
_{+},\mathcal{L}_{-}\right] &=&0  \notag \\
\partial _{\bar{\zeta}}\mathcal{L}_{+}-\partial _{+}\mathcal{L}_{\bar{\zeta}%
}+\left[ \mathcal{L}_{\bar{\zeta}},\mathcal{L}_{+}\right] &=&0  \label{lll}
\\
\partial _{\bar{\zeta}}\mathcal{L}_{-}-\partial _{-}\mathcal{L}_{\bar{\zeta}%
}+\left[ \mathcal{L}_{\bar{\zeta}},\mathcal{L}_{-}\right] &=&0  \notag
\end{eqnarray}%
involving three pairs of generalised Lax pairs $\left( \mathcal{L}_{+},%
\mathcal{L}_{-}\right) $, $\left( \mathcal{L}_{\bar{\zeta}},\mathcal{L}%
_{+}\right) $ and $\left( \mathcal{L}_{\bar{\zeta}},\mathcal{L}_{-}\right) $
with coordinate dependence as 
\begin{equation}
\begin{tabular}{lll}
$\mathcal{L}_{+}$ & $=$ & $\mathcal{L}_{+}\left( X^{+},X^{-};\zeta ,\bar{%
\zeta};t_{a}\right) $ \\ 
$\mathcal{L}_{-}$ & $=$ & $\mathcal{L}_{-}\left( X^{+},X^{-};\zeta ,\bar{%
\zeta};t_{a}\right) $ \\ 
$\mathcal{L}_{\bar{\zeta}}$ & $=$ & $\mathcal{L}_{\bar{\zeta}}\left(
X^{+},X^{-};\zeta ,\bar{\zeta};t_{a}\right) $%
\end{tabular}
\label{lp}
\end{equation}%
Roughly speaking, these three extended Lax pairs $\left( \mathcal{L}_{+},%
\mathcal{L}_{-}\right) $, $\left( \mathcal{L}_{\bar{\zeta}},\mathcal{L}%
_{+}\right) $ and $\left( \mathcal{L}_{\bar{\zeta}},\mathcal{L}_{-}\right) $
could be interpreted as obeying Lax- type equation in the 2d subspaces $%
\left( X^{+},X^{-}\right) ,\left( X^{+},\bar{\zeta}\right) $ and $\left(
X^{-},\bar{\zeta}\right) $. A field realisation of the $\mathcal{L}_{\pm }$
and $\mathcal{L}_{\bar{\zeta}}$ operators extending the 2d realisation (\ref%
{4}) can be obtained by thinking of the 4d $\mathcal{L}_{\pm }$ as related
to the 2d $L_{\pm }$ like%
\begin{equation}
L_{+}=\left. \mathcal{L}_{+}\right \vert _{\zeta =\bar{\zeta}=0}\qquad
,\qquad L_{-}=\left. \mathcal{L}_{-}\right \vert _{\zeta =\bar{\zeta}=0}
\label{lm}
\end{equation}%
This feature suggests that $\mathcal{L}_{\pm }$ and $\mathcal{L}_{\bar{\zeta}%
}$ can be realised in terms of the following field system%
\begin{equation}
\begin{tabular}{lll}
$\Phi $ & $=$ & $\Phi \left( X^{+},X^{-};\zeta ,\bar{\zeta}\right) $ \\ 
$\Psi $ & $=$ & $\Psi _{-}^{+}\left( X^{-},\zeta ,\bar{\zeta}\right) $ \\ 
$N$ & $=$ & $N_{+}^{-}\left( X^{+},\zeta ,\bar{\zeta}\right) $ \\ 
$\Gamma $ & $=$ & $\Gamma _{\bar{\zeta}}^{+}\left( \zeta ,\bar{\zeta}\right) 
$%
\end{tabular}
\label{39}
\end{equation}%
extending the 2d system $\left \{ \phi ,\alpha ,\beta \right \} $ used in
the 2d Lax construction (\ref{4}) with $\phi =\phi \left( X^{+},X^{-}\right) 
$ but $\alpha $ and $\beta $ constant parameters; $\frac{\partial \alpha }{%
\partial X^{\pm }}=\frac{\partial \beta }{\partial X^{\pm }}=0$. General
arguments indicate that $\mathcal{L}_{\pm }$ and $\mathcal{L}_{\bar{\zeta}}$
are realised like%
\begin{eqnarray}
\mathcal{L}_{+} &=&\frac{\partial \Phi }{\partial X^{+}}h-NE^{+}  \notag \\
\mathcal{L}_{-} &=&e^{2\Phi }\Psi E^{-}  \label{40} \\
\mathcal{L}_{\bar{\zeta}} &=&-\frac{\partial \log N}{2\partial \bar{\zeta}}%
h+\Gamma e^{2\Phi }E^{-}  \notag
\end{eqnarray}%
The charges carried by the fields in (\ref{39}) are as before and are of two
types: space and Lie algebraic; they should be understood like for instance $%
\mathcal{L}_{+}=\frac{\partial \Phi }{\partial X^{+}}h-N_{+}^{-}E^{+},$ $%
\mathcal{L}_{-}=e^{2\Phi }\Psi _{-}^{+}E^{-}$\ and so on; for convenience
the charges of $N_{+}^{-}$, $\Psi _{-}^{+}$ and $\Gamma _{-}^{+}$ have been
omitted in (\ref{40}). Moreover, the dependence of the fields $\left \{ \Phi
,N,\Psi ,\Gamma \right \} $ into the variables $X^{+},X^{-}$ and $\zeta ,%
\bar{\zeta}$ as specified in (\ref{39}) is motivated by the fact that we
want to reproduce the Liouville equation; for instance when computing $%
\partial _{-}\mathcal{L}_{+}$ we need $\frac{\partial N}{\partial X^{-}}=0$
requiring that $N$ should not depend on $X^{-}$. By substituting (\ref{40})
back into eqs(\ref{lll}) and using the commutation relations of sl$\left(
2\right) $, we obtain the three following equations 
\begin{eqnarray}
\left( \frac{\partial ^{2}\Phi }{\partial X^{-}\partial X^{+}}+N\Psi
e^{2\Phi }\right) h &=&0  \label{ll0} \\
\left( \frac{\partial ^{2}\Phi }{\partial \bar{\zeta}\partial X^{+}}+N\Gamma
e^{2\Phi }\right) h &=&0  \label{ll1} \\
\left( \frac{\partial \Psi }{\partial \bar{\zeta}}+\frac{\Psi }{N}\frac{%
\partial N}{\partial \bar{\zeta}}+2\Psi \frac{\partial \Phi }{\partial \bar{%
\zeta}}-2\Gamma \frac{\partial \Phi }{\partial X^{-}}\right) E^{+} &=&0
\label{ll2}
\end{eqnarray}%
Two of these relations point in the diagonal h- direction of sl$\left(
2\right) $, and look like generalisations of the standard Liouville
equation; the third relation points in E$^{+}$- direction and behaves as a
constraint relation between the fields (\ref{39}).

\subsubsection{More on eqs (\protect \ref{ll0}-\protect \ref{ll2})}

The above relations (\ref{ll0}-\ref{ll2}) obey some special features
capturing data on the 2d Liouville equation among which the two following
ones; other properties like exact solution will be given in next
sub-subsection:

\  \  \ 

$\left( 1\right) $ \emph{Symmetry under holomorphic change on} $\mathcal{C}$%
. \newline
Eqs(\ref{ll0}-\ref{ll2}) are invariant under the holomorphic local change 
\begin{eqnarray}
\Psi \qquad &\rightarrow &\qquad e^{\mathbf{f}\left( \zeta \right) }\times
\Psi  \notag \\
N\qquad &\rightarrow &\qquad e^{-\mathbf{k}\left( \zeta \right) }\times N 
\notag \\
\Gamma \qquad &\rightarrow &\qquad e^{\mathbf{f}\left( \zeta \right) }\times
\Gamma  \label{t1} \\
\Phi \qquad &\rightarrow &\qquad \Phi +\frac{1}{2}\mathbf{k}\left( \zeta
\right) -\frac{1}{2}\mathbf{f}\left( \zeta \right)  \notag
\end{eqnarray}%
where $\mathbf{f}\left( \zeta \right) =\mathbf{f}$ and $\mathbf{k}\left(
\zeta \right) =\mathbf{k}$ are arbitrary holomorphic functions in the
complex coordinate $\zeta $ of the base space $\mathcal{C}$. Under these
transformations, we also have%
\begin{eqnarray}
\frac{\partial \Psi }{\partial \bar{\zeta}}\qquad &\rightarrow &\qquad e^{%
\mathbf{f}}\times \frac{\partial \Psi }{\partial \bar{\zeta}}  \notag \\
\frac{\partial N}{\partial \bar{\zeta}}\qquad &\rightarrow &\qquad e^{-%
\mathbf{k}}\times \frac{\partial N}{\partial \bar{\zeta}}  \notag \\
\frac{\partial \Gamma }{\partial \bar{\zeta}}\qquad &\rightarrow &\qquad e^{%
\mathbf{f}}\times \frac{\partial \Gamma }{\partial \bar{\zeta}}  \label{t2}
\\
\frac{\partial \Phi }{\partial \bar{\zeta}}\qquad &\rightarrow &\qquad \frac{%
\partial \Phi }{\partial \bar{\zeta}}  \notag
\end{eqnarray}%
Observe moreover that if choosing $\mathbf{f}=\mathbf{k}$, then the 4d
scalar field $\Phi $\ becomes invariant; and by thinking of $N$ and $\Psi $
as real quantities like%
\begin{equation}
\Psi =\beta e^{\mathbf{f}\left( \zeta \right) +\mathbf{\bar{f}}\left( \bar{%
\zeta}\right) }\qquad ,\qquad N=\frac{\alpha }{e^{\mathbf{k}\left( \zeta
\right) +\mathbf{\bar{k}}\left( \bar{\zeta}\right) }}\qquad ,\qquad \Gamma
=\gamma e^{\mathbf{f}\left( \zeta \right) +\mathbf{\bar{f}}\left( \bar{\zeta}%
\right) }  \label{rc}
\end{equation}%
where $\alpha $ and $\beta $ are as above, then the product 
\begin{equation}
N\Psi =\alpha \beta \times \frac{e^{\mathbf{f}}}{e^{\mathbf{k}}}\times \frac{%
e^{\mathbf{\bar{f}}}}{e^{\mathbf{\bar{k}}}}  \label{fk}
\end{equation}%
behaves as%
\begin{equation}
N\Psi =\alpha \beta \times \left \vert \frac{e^{\mathbf{f}}}{e^{\mathbf{k}}}%
\right \vert ^{2}\qquad ,\qquad N\Gamma =\alpha \gamma \times \left \vert 
\frac{e^{\mathbf{f}}}{e^{\mathbf{k}}}\right \vert ^{2}  \label{a}
\end{equation}%
and reduced further to $N\Psi =\alpha \beta $\  \textrm{and} $N\Gamma =\alpha
\gamma $ if we take $\mathbf{f}=\mathbf{k}$.

\  \  \  \ 

$\left( 2\right) $ \emph{From (\ref{ll0}-\ref{ll2}) to Liouville equation.}%
\newline
The formal similarity between the two relations (\ref{ll0}) and (\ref{ll1})
is striking an suggestive since both of them describe an extension of the 2d
Liouville equation. However, these two relations can \textrm{be brought to
one equation namely} 
\begin{equation}
\frac{\partial ^{2}\Phi }{\partial X^{-}\partial X^{+}}+\alpha \beta
e^{2\Phi }=0  \label{ab}
\end{equation}%
if we think of the third (\ref{ll2}) like%
\begin{equation}
\left( \frac{\partial \Psi }{\partial \bar{\zeta}}+\frac{\Psi }{N}\frac{%
\partial N}{\partial \bar{\zeta}}\right) +2\left( \Psi \frac{\partial \Phi }{%
\partial \bar{\zeta}}-\Gamma \frac{\partial \Phi }{\partial X^{-}}\right) =0
\end{equation}%
and cast it as follows%
\begin{eqnarray}
\frac{\partial \Psi }{\partial \bar{\zeta}}+\frac{\Psi }{N}\frac{\partial N}{%
\partial \bar{\zeta}} &=&0  \notag \\
\Psi \frac{\partial \Phi }{\partial \bar{\zeta}}-\Gamma \frac{\partial \Phi 
}{\partial X^{-}} &=&0  \label{np}
\end{eqnarray}%
or equivalently%
\begin{equation}
\frac{\partial }{\partial \bar{\zeta}}\log \left( N\Psi \right) =0\qquad
,\qquad \frac{\partial \Phi }{\partial \bar{\zeta}}=\frac{\Gamma }{\Psi }%
\frac{\partial \Phi }{\partial x^{-}}
\end{equation}%
The first relation shows that $\log \left( N\Psi \right) $ is independent of 
$\bar{\zeta}$ and $\zeta $; and then $N\times \Psi $ is a constant precisely
given by the Liouville coupling constant $\kappa =\alpha \times \beta $ as
one sees from (\ref{a}). The other relation in (\ref{np}) namely $\frac{%
\partial \Phi }{\partial \bar{\zeta}}=\frac{\Gamma }{\Psi }\frac{\partial
\Phi }{\partial X^{-}}$ is also remarkable in the sense it relates $\frac{%
\partial \Phi }{\partial \bar{\zeta}}$ to the gradient $\frac{\partial \Phi 
}{\partial X^{-}}$. By substituting it back into eq(\ref{ll1}), we obtain 
\begin{equation}
\frac{\Gamma }{\Psi }\frac{\partial ^{2}\Phi }{\partial X^{-}\partial X^{+}}%
+N\Gamma e^{2\Phi }=0
\end{equation}%
Multiplying with $\frac{\Psi }{\Gamma }$ and using $N\Psi =\kappa $, we
obtain $\frac{\partial ^{2}\Phi _{\zeta }}{\partial X^{-}\partial X^{+}}%
+\kappa e^{2\Phi _{\zeta }}$ which is exactly with the generalised Liouville
equation (\ref{ab}) with $\Phi _{\zeta }=\Phi \left( X^{+},X^{-};\bar{\zeta}%
,\zeta \right) $.

\subsubsection{Exact solution of (\protect \ref{ll0}-\protect \ref{ll2})}

To work out the exact solution of eqs(\ref{ll0}-\ref{ll2}), we use results
from 2d Liouville theory and its conformal symmetry. To that purpose, we
shall solve each of the eqs (\ref{ll0}-\ref{ll2}) separately by starting
from (\ref{ll0}), then solving (\ref{ll1}) and finally solving (\ref{ll2}).
First, notice that (\ref{ll0}) looks like a generalised Liouville equation
in the plane $\left( X^{+},X^{-}\right) $; but with coupling given by $%
K_{+-}=N\Psi ,$%
\begin{equation}
\frac{\partial ^{2}\Phi }{\partial X^{-}\partial X^{+}}+K_{+-}e^{2\Phi }=0
\label{522}
\end{equation}%
Comparing this equation with (\ref{s1}), one can easily wonder an exact
solution of (\ref{522}) in terms of the variables $X^{\pm }$ and the
coupling $K_{+-}$. The solution has a similar structure as eq(\ref{fd}) and
reads, up to a conformal transformation, \textrm{as follows}%
\begin{equation}
\Phi =\ln \left( \frac{1}{1+K_{+-}X^{+}X^{-}}\right)  \label{s01}
\end{equation}%
with $K_{+-}$ having no dependence on the $X^{\pm }$- variables; i.e:%
\begin{equation}
K_{+-}=K\left( \zeta ,\bar{\zeta}\right)
\end{equation}%
Viewed from the 2d Liouville side, this solution corresponds to a fibration
of (\ref{fd}) on the complex curve $\mathcal{C}$. Notice also the following
expression to be encountered later on%
\begin{equation}
X^{-}\frac{\partial \Phi }{\partial X^{-}}=-\frac{K_{+-}X^{+}X^{-}}{%
1+K_{+-}X^{-}X^{+}}  \label{f1}
\end{equation}%
Regarding the second relation (\ref{ll1}), it looks as well like a Liouville
type equation; but in the plane $\left( X^{+},\bar{\zeta}\right) ,$ and a
different coupling given by $K_{+\bar{\zeta}}=N\Gamma $, 
\begin{equation}
\frac{\partial ^{2}\Phi }{\partial \bar{\zeta}\partial X^{+}}+K_{+\bar{\zeta}%
}e^{2\Phi }=0
\end{equation}%
An exact solution of this equation is derived easily by using the same trick
as above; it reads, up to a conformal transformation, like%
\begin{equation}
\Phi =\ln \left( \frac{1}{1+K_{+\bar{\zeta}}X^{+}\bar{\zeta}}\right)
\label{s02}
\end{equation}%
with $K_{+\bar{\zeta}}$ having no dependence on the variables $X^{+}$ and $%
\bar{\zeta}$; i.e: 
\begin{equation}
K_{+\bar{\zeta}}=\tilde{K}\left( X^{-},\zeta \right)
\end{equation}%
By equating the two solutions (\ref{s01}) and (\ref{s02}) as they concern
the same field $\Phi $, we end with the identification $K_{+\bar{\zeta}}X^{+}%
\bar{\zeta}=K_{+-}X^{+}X^{-}$ leading to 
\begin{equation}
K_{+\bar{\zeta}}\bar{\zeta}=K_{+-}X^{-}\qquad \Rightarrow \qquad \Gamma \bar{%
\zeta}=\Psi X^{-}
\end{equation}%
and implying in turns%
\begin{equation}
\frac{K_{+\bar{\zeta}}}{K_{+-}}=\frac{\Gamma }{\Psi }=\frac{X^{-}}{\bar{\zeta%
}}
\end{equation}%
Notice that the equality $K_{+\bar{\zeta}}X^{+}\bar{\zeta}=K_{+-}X^{+}X^{-}$
shows that we also have 
\begin{equation}
\bar{\zeta}\frac{\partial \Phi }{\partial \bar{\zeta}}=X^{-}\frac{\partial
\Phi }{\partial X^{-}}  \label{id}
\end{equation}%
To determine the solution of the third relation (\ref{ll2}), it is
interesting to put it into a convenient form. By dividing (\ref{ll2}) by $%
\Psi $, we get 
\begin{equation}
\frac{\partial \log \Psi }{\partial \bar{\zeta}}+\frac{\partial \log N}{%
\partial \bar{\zeta}}+2\frac{\partial \Phi }{\partial \bar{\zeta}}-2\frac{%
\Gamma }{\Psi }\frac{\partial \Phi }{\partial X^{-}}=0
\end{equation}%
Then by substituting $\frac{\Gamma }{\Psi }=\frac{X^{-}}{\bar{\zeta}}$, we
can put it into the form%
\begin{equation}
\bar{\zeta}\frac{\partial \log \left( N\Psi \right) }{\partial \bar{\zeta}}+2%
\bar{\zeta}\frac{\partial \Phi }{\partial \bar{\zeta}}-2X^{-}\frac{\partial
\Phi }{\partial X^{-}}=0
\end{equation}%
But because of the identity (\ref{id}), we end with%
\begin{equation}
\bar{\zeta}\frac{\partial \log \left( N\Psi \right) }{\partial \bar{\zeta}}=0
\end{equation}%
\textrm{showing that} $N\Psi $ is independent from $\bar{\zeta}$.

\subsection{Quantum line operators}

In this subsection, we build line operators associated with the pair $\left( 
\mathcal{L}_{+},\mathcal{L}_{-}\right) $ extending the usual Lax pair $%
\left( L_{+},L_{-}\right) $ linearising the 2d Liouville equation. The
structure of these operators are given by eqs(\ref{wr}) and (\ref{19}); but
with the partial gauge connection $\mathcal{A}=\mathcal{A}_{\mu }dX^{\mu }$
replaced by a constrained gauge connection $\mathcal{B}=\mathcal{B}_{\mu
}dX^{\mu }$ related to $\mathcal{A}$ like 
\begin{equation}
\mathcal{B}=dX^{\mu }\Gamma _{\mu }^{\rho }\mathcal{A}_{\rho }  \label{ba}
\end{equation}%
\textrm{where} the operator $\mathbf{\Gamma }_{\mu }^{\rho }=t_{b}\Gamma
_{\mu }^{\rho b}$ will be determined below. To get the explicit of $\Gamma
_{\mu }^{\rho }\mathcal{A}_{\rho }$, we first need to determine the analogue
of the constraint eqs(\ref{24}) to the 4d space $\mathbb{M}_{4}=\Sigma
\times \mathcal{C}$.

\subsubsection{Deriving the constraint equations}

To begin, notice that the realisation (\ref{40}) of the $\mathcal{L}_{\pm },%
\mathcal{L}_{\bar{\zeta}}$ operators in terms of the fields (\ref{39}) can
be rigorously derived by solving the vanishing condition of the\ CWY
curvature 
\begin{equation}
\mathcal{F}_{\mu \nu }=\partial _{\mu }\mathcal{A}_{\nu }-\partial _{\nu }%
\mathcal{A}_{\mu }+\left[ \mathcal{A}_{\mu },\mathcal{A}_{\nu }\right]
\end{equation}%
by using the same method as the one used in sub-subsection 4.2.1 for the 2d
Liouville theory: see the analysis between eq(\ref{1a}) and eq(\ref{1b}).
This solution is obtained by imposing constraints on some components of the
non abelian vector potential $\mathcal{A}_{\mu }^{a}$ as done for (\ref{24}%
). Recall that $\mathcal{A}=\mathcal{A}_{\mu }dX^{\mu }$ is a partial gauge
connection valued in sl$\left( 2\right) $ with 3d subspace index as $\mu
=+,-,\bar{\zeta}$ and metric $\eta _{\mu \nu }$ like 
\begin{equation}
\eta _{\mu \nu }=\left( 
\begin{array}{ccc}
0 & 1 & 0 \\ 
1 & 0 & 0 \\ 
0 & 0 & \frac{1}{2}%
\end{array}%
\right) \qquad ,\qquad \eta ^{\mu \nu }=\left( 
\begin{array}{ccc}
0 & 1 & 0 \\ 
1 & 0 & 0 \\ 
0 & 0 & 2%
\end{array}%
\right)  \label{me}
\end{equation}%
Each one of the $\mathcal{A}_{\mu }$ components expands along the sl$\left(
2\right) $ generators as $t_{a}\mathcal{A}_{\mu }^{a}$ with $t_{1},$ $t_{2},$
$t_{3}$ \ standing for the generators of sl$\left( 2\right) $. To exhibit
the constraint eqs on the $\mathcal{A}_{\mu }^{a}$ that lead to (\ref{40}),
we use the Cartan basis $\left( h,E^{\pm }\right) $ of sl$\left( 2\right) $
and expand the non abelian vector potential$\mathcal{\ }$as follows 
\begin{equation}
\begin{tabular}{lll}
$\mathcal{A}_{+}$ & $=$ & $\mathcal{A}_{+}^{0}h+\mathcal{A}_{+}^{-}E^{+}+%
\mathcal{A}_{+}^{+}E^{-}$ \\ 
$\mathcal{A}_{-}$ & $=$ & $\mathcal{A}_{-}^{0}h+\mathcal{A}_{-}^{-}E^{+}+%
\mathcal{A}_{-}^{+}E^{-}$ \\ 
$\mathcal{A}_{\bar{\zeta}}$ & $=$ & $\mathcal{A}_{\bar{\zeta}}^{0}h+\mathcal{%
A}_{\bar{\zeta}}^{-}E^{+}+\mathcal{A}_{\bar{\zeta}}^{+}E^{-}$%
\end{tabular}
\label{41}
\end{equation}%
Generally speaking, the above equations teach us that in the sl$\left(
2\right) $ case, the non abelian vector potential has nine components $%
\mathcal{A}_{\mu }^{a}$ since $\mu =\pm ,\bar{\zeta}$ and $a=0,\pm $. By
comparing the above $\mathcal{A}_{0,\pm }$- expansions with those of the $%
\mathcal{L}_{0,\pm }$'s given by eqs (\ref{40}) that we rewrite like, 
\begin{eqnarray}
\mathcal{L}_{+} &=&\frac{\partial \Phi }{\partial X^{+}}h-NE^{+}+0E^{-} 
\notag \\
\mathcal{L}_{-} &=&\text{\  \ }0\text{ }h\text{ \  \ }+\text{ \ }%
0E^{+}+e^{2\Phi }\Psi E^{-} \\
\mathcal{L}_{\bar{\zeta}} &=&-\frac{h}{2N}\frac{\partial N}{\partial \bar{%
\zeta}}\text{ }+0\text{ }E^{+}+\Gamma e^{2\Phi }E^{-}  \notag
\end{eqnarray}%
we deduce the constraint relations that we have to impose some of the $%
\mathcal{A}_{\mu }^{0,\pm }$'s in order to embed the Liouville equation into
the CWY theory. These constraints are given by 
\begin{equation}
\mathcal{A}_{+}^{+}=0\qquad ,\qquad \mathcal{A}_{-}^{0}=0\qquad ,\qquad 
\mathcal{A}_{-}^{-}=0\qquad ,\qquad \mathcal{A}_{\bar{\zeta}}^{-}=0
\label{523}
\end{equation}%
leaving then five non zero components filling particular directions in the
gauge bundle. The above constraints extend (\ref{24}); their gauge
invariance requires reducing down the volume of the SL$\left( 2\right) $ set
of gauge transformations on $\mathbb{M}_{4}=\Sigma \times \mathcal{C}$. This
is because the generic SL$\left( 2\right) $ change 
\begin{equation}
\mathcal{A}_{\mu }^{\prime }=g^{-1}\mathcal{A}_{\mu }g+g^{-1}\partial _{\mu
}g
\end{equation}%
does not preserve (\ref{523}). The above constraints require also reducing
down the volume of the Diff$\left( \Sigma \right) $ set of general
coordinate transformations $\tilde{X}^{\pm }=f^{\pm }\left(
X^{+},X^{-}\right) $. For the gauge symmetry, we have in addition to (\ref%
{eeh}), the extra condition along the complex $\mathcal{C}$ curve dimension%
\begin{equation}
Tr\left( E^{-}g^{-1}\frac{\partial }{\partial \bar{\zeta}}g\right) =0\quad
,\quad g=e^{\vartheta }\quad ,\quad \vartheta =\vartheta ^{0}h+\vartheta
^{-}E^{+}+\vartheta ^{+}E^{-}  \label{he}
\end{equation}%
with $\vartheta $ a priori a generic function of the coordinates $\left(
X^{+},X^{-},\zeta ,\bar{\zeta}\right) $. The conditions (\ref{eeh}) and (\ref%
{he}) on the local matrix transformations $g=e^{\vartheta }$ can be solved
as before with analytic gauge parameters $\vartheta ^{0}=\vartheta
^{0}\left( X^{+},\zeta \right) $ and $\vartheta ^{-}=\vartheta ^{-}\left(
X^{+},\zeta \right) $. The same thing holds for the set of general
coordinate transformations which gets reduced to invariance under
holomorphic transformations $\tilde{X}^{+}=f\left( X^{+}\right) $ and $%
\tilde{X}^{-}=\bar{f}\left( X^{-}\right) $.

\subsubsection{Building line operators}

To build line operators associated with the embedding of Liouville equation
in the CWY theory, we use the eqs(\ref{wr}) and (\ref{19}) and impose the
constraints in the holonomy like%
\begin{equation}
\left. 
\begin{array}{c}
\text{ \ } \\ 
\mathbf{\varphi }\left( K_{\zeta }\right) \\ 
\text{ \  \ }%
\end{array}%
\right \vert _{\substack{ \text{ \  \ }  \\ Eq({\small \ref{523}})}}=\left(
\doint \nolimits_{K_{_{\zeta }}}\mathcal{A}\right) _{Eq({\small \ref{523}})}
\end{equation}%
Substituting the constraint eqs (\ref{523}) back into (\ref{41}), one gets a
restriction on the allowed directions of the vector potentials $\mathcal{A}%
_{\mu }^{a}$ in the $SL\left( 2\right) $ gauge bundle. Thinking of these
restricted directions in terms of \textrm{projections }in $sl\left( 2\right) 
$, we can define the constrained vector potentials like $\mathcal{B}_{\mu
}^{b}$ related to the generic $\mathcal{A}_{\mu }^{a}$ as follows 
\begin{equation}
\mathcal{B}_{\mu }^{b}=\Gamma _{\mu a}^{\rho b}\mathcal{A}_{\rho }^{a}
\label{LC}
\end{equation}%
The tensor $\Gamma _{\mu a}^{\rho b}$ which links the projected $\mathcal{B}%
_{\mu }^{b}$ to the generic $\mathcal{A}_{\rho }^{a}$ can be presented in
different, but equivalent, manners depending on the index we want to
exhibit; that is the space indices $\mu ,\rho $;\ or the sl$\left( 2\right) $
Lie algebra ones $a,b$. For instance, by multiplying both sides of (\ref{LC}%
) by $t_{b}$ generator, we can put the above relation into the matrix form $%
\mathcal{B}_{\mu }=\boldsymbol{\Gamma }_{\mu a}^{\rho }\mathcal{A}_{\rho
}^{a}$ where now we have $\boldsymbol{\Gamma }_{\mu a}^{\rho }=t_{b}\Gamma
_{\mu a}^{\rho b}$. Moreover, using the \textrm{Killing form} normalised
like $Tr\left( t_{a}t_{b}\right) =\delta _{ab}$, we can express the above
relation in terms of $\mathcal{B}_{\mu }=t_{b}\mathcal{B}_{\mu }^{b}$ and $%
\mathcal{A}_{\rho }=t_{a}\mathcal{A}_{\rho }^{a}$ as follows 
\begin{equation}
\mathcal{B}_{\mu }=\mathbf{\Gamma }_{\mu }^{\rho }\mathcal{A}_{\rho }
\end{equation}%
where $\mathbf{\Gamma }_{\mu a}^{\rho }=\mathbf{\Gamma }_{\mu }^{\rho
}\left( t_{a}\right) $. In the Cartan basis of sl$\left( 2\right) $ with
generators $\left( h,E^{\pm }\right) $, the explicit expression of $\Gamma
_{\mu a}^{\rho b}$ can be read from the following expansions of $\mathbf{%
\Gamma }_{\mu a}^{\rho },$%
\begin{equation}
\begin{tabular}{lll}
$\mathbf{\Gamma }_{+a}^{\rho }$ & $=$ & $\eta _{+}^{\rho }\left[ \frac{h}{2}%
Tr\left( ht_{a}\right) +E^{+}Tr\left( E^{-}t_{a}\right) \right] $ \\ 
$\mathbf{\Gamma }_{-a}^{\rho }$ & $=$ & $\eta _{-}^{\rho }E^{-}Tr\left(
E^{+}t_{a}\right) $ \\ 
$\mathbf{\Gamma }_{\bar{\zeta}a}^{\rho }$ & $=$ & $\eta _{\bar{\zeta}}^{\rho
}\left[ \frac{h}{2}Tr\left( ht_{a}\right) +E^{-}Tr\left( E^{+}t_{a}\right) %
\right] $%
\end{tabular}
\label{gg}
\end{equation}%
where $\eta _{\mu \nu }$ is the 3d metric (\ref{me}) \textrm{and where we
have used }$(\eta _{\bar{\zeta}}^{\rho })^{2}=\eta _{\bar{\zeta}}^{\rho }$.
For example, we have $\Gamma _{+0}^{\rho 0}=\eta _{+}^{\rho }$, $\Gamma
_{-0}^{\rho 0}=0$ and $\Gamma _{\bar{\zeta}0}^{\rho 0}=\eta _{\bar{\zeta}%
}^{\rho }$ with $\eta _{\mu \rho }\eta ^{\rho \nu }=\eta _{\mu }^{\rho }$
refering to the Kronecker symbol $\delta _{\mu }^{\rho }$.\newline
With the gauge connection $\mathcal{B}$ at hand, we can construct line
operators associated with Liouville field by mimicking the derivation of eqs(%
\ref{wr}) and (\ref{19}). Observables associated with the holonomy of $%
\mathcal{B}$ are then given by Wilson like operators and their
generalisations introduced in section 2. As an example, we have, 
\begin{equation}
W_{\varrho _{sl_{2}}}\left[ K_{\zeta }\right] =Tr_{\varrho _{sl_{2}}}\left[
P\exp \left( \int_{K_{\zeta }}\mathcal{B}\right) \right]  \label{b}
\end{equation}%
where $\varrho _{sl_{2}}$ refer to some representation of sl$\left( 2\right) 
$ and the curve $K_{\zeta }$ lies in the topological plane $\Sigma $ and is
defined as in the CWY theory of section 2. Notice that instead of the usual
generic partial gauge $\mathcal{A}=dX^{\mu }\mathcal{A}_{\mu }^{a}$ we have
now the constrained gauge connection $\mathcal{B}=\mathcal{B}_{\mu }dX^{\mu
} $ which is related to the generic $\mathcal{A}$ through (\ref{LC}).
Denoting by $\mathbf{\psi }_{K_{\zeta }}=t_{b}\psi _{K_{\zeta }}^{b}$\ the
holonomy of $\mathcal{B}$ along the curve $K_{\zeta }$, we then have%
\begin{equation}
\mathbf{\psi }_{K_{\zeta }}^{b}=\int_{K_{\zeta }}\Gamma _{\mu a}^{\nu b}%
\mathcal{A}_{\nu }^{a}dX^{\mu }  \label{bd}
\end{equation}%
Moreover, because of the fact that $K_{\zeta }$ belongs to the topological
plane $\Sigma $ taken here as $\mathbb{R}^{2}$, it is clear that the
contribution to the holonomy is given by 
\begin{equation}
\mathbf{\psi }_{K_{\zeta }}^{b}=\int_{K_{\zeta }}\mathcal{A}_{\nu
}^{a}\left( \Gamma _{+a}^{\nu b}dX^{+}+\Gamma _{-a}^{\nu b}dX^{-}\right)
\end{equation}%
Notice also that, in addition to the coordinate variables $\left( x,y\right) 
$ of $\mathbb{R}^{2}$, the $\mathcal{B}_{\mu }^{a}$ depends as well on the
spectral parameter $\zeta $. Using the same trick as done in (\ref{xp}) by
keeping only the holomorphic variable $\zeta $, we\ obtain a new operator $%
\boldsymbol{\hat{B}}_{\pm }\left( x,y,\zeta \right) $ defined as 
\begin{equation}
\boldsymbol{\hat{B}}_{\pm }\left( x,y,\zeta \right) :=\sum_{k=0}^{\infty
}t_{a,n}\hat{B}_{\pm }^{a\left( n\right) }\left( x,y\right)
\end{equation}%
with extended generators $t_{a,n}=t_{a}\otimes \zeta ^{n}.$ By using these
quantities, we can construct generalised like operators for Liouville theory
like 
\begin{equation}
\hat{W}_{\mathbf{\hat{\varrho}}_{sl_{2}}}\left[ K_{\zeta }\right] =Tr_{%
\mathbf{\hat{\varrho}}_{sl_{2}}}\left( P\exp (\int_{K_{\zeta }}\boldsymbol{%
\hat{B}}_{+}dX^{+}+\boldsymbol{\hat{B}}_{-}dX^{-})\right)
\end{equation}%
where now $\mathbf{\hat{\varrho}}_{sl_{2}}$ is a representation of the
infinite dimensional algebras $sl_{2}\left[ \left[ \zeta \right] \right] $
induced by the fibration of sl$\left( 2\right) $ on the holomorphic complex
line $\mathbb{C}$. Similar comments that have been done for the derivation
of eq(\ref{20}) applies as well here for the above $\hat{W}_{\mathbf{\hat{%
\varrho}}_{sl_{2}}}\left[ K_{\zeta }\right] $.

\section{One loop quantum effect}

\begin{figure}[tbph]
\begin{center}
\hspace{0cm} \includegraphics[width=10cm]{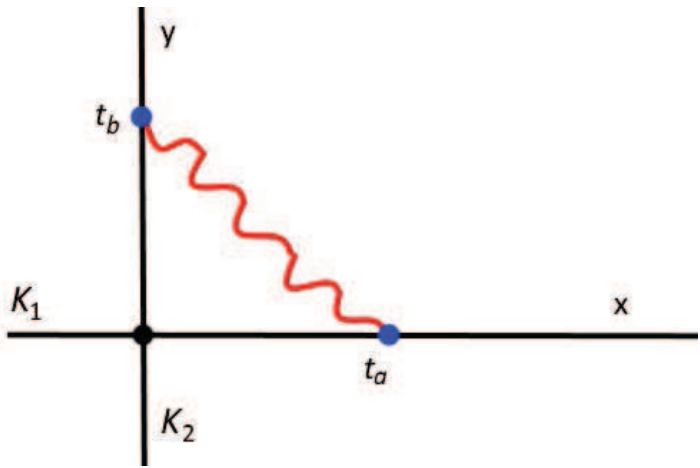}
\end{center}
\par
\vspace{-0.5 cm}
\caption{One gluon exchange between two Wilson lines operators $W_{\hat{%
\protect \varrho}_{1}}\left( K_{1}\right) $ and $W_{\hat{\protect \varrho}%
_{2}}\left( K_{2}\right) $ supported by $K_{1}$ and $K_{2}$ with spectral
parameters $\protect \zeta _{1}$ and $\protect \zeta _{2}$. These line
operators are respectively chosen as given by the x- and y- axes in the
toplogical plane $\mathbb{R}^{2}$.}
\label{3a}
\end{figure}

In this section, we calculate the expression of the amplitude of two
intersecting lines $K_{\zeta _{1}}$ and $K_{\zeta _{2}}$ supporting two
quantum Wilson operators $W_{\hat{\varrho}_{1}}\left[ K_{\zeta _{1}}\right] $
and $W_{\hat{\varrho}_{2}}\left[ K_{\zeta _{2}}\right] $ as schematised in
the figure \ref{3a}. Then, we compare the obtained result with a similar
amplitude calculated in \cite{1F} for generic vector potentials. 
The two quantum line operators of the figure \ref{3a} are based on
holonomies of type (\ref{bd}); they exchange one gluon and form altogether a
one- loop Feynman diagram. The two Wilson operators are respectively
characterised by the spectral parameters $\zeta _{1}$ and $\zeta _{2}$; and
carry $\varrho _{1}$ and $\varrho _{2}$\ representations of sl$\left(
2\right) $. The amplitude $\mathcal{I}_{1}=\mathcal{I}_{1}\left[ \varrho
_{1},\zeta _{1};\varrho _{2},\zeta _{2}\right] $ of this one- loop Feynman
diagram is proportional to $\hbar $ and, because of translation invariance
in the 4d space $\mathbb{M}_{4}=\mathbb{C}\times \mathbb{C}$, is a function
of $\zeta _{1}-\zeta _{2}$; so the $\mathcal{I}_{1}$ has the form%
\begin{equation}
\mathcal{I}_{1}=\hbar F_{\varrho _{1},\varrho _{2}}\left( \zeta _{1}-\zeta
_{2}\right)
\end{equation}%
where $F_{\varrho _{1},\varrho _{2}}\left( \zeta _{1}-\zeta _{2}\right) $ is
obtained by computing the contribution of the diagram by using the Feynman
rules given in section 2. The calculation of $\mathcal{I}_{1}$ follows the
same manner as done in \textrm{\cite{1F}}\ for a generic non abelian gauge
potential $\mathcal{A}_{\mu }$. The main difference is that now the two
quantum lines $W_{\hat{\varrho}_{1}}\left[ K_{\zeta _{1}}\right] $ and $W_{%
\hat{\varrho}_{2}}\left[ K_{\zeta _{2}}\right] $ are built out of the vector
potential $\mathcal{B}_{\mu }^{a}$ which is related to the CWY gauge field $%
\mathcal{A}_{\mu }$ like in (\ref{LC}). In what follows, we work out this
computation for the one- loop diagram \ref{3a} built out of the vector
potential $\mathcal{B}_{\mu }$; and show that the quantum contribution has
the form 
\begin{equation}
\mathcal{I}_{1}=\hbar _{\text{ }}\frac{\tilde{c}_{\varrho _{1},\varrho _{2}}%
}{\zeta _{1}-\zeta _{2}}  \label{i1}
\end{equation}%
with coefficient $\tilde{c}_{\varrho ,\varrho ^{\prime }}$ given by%
\begin{equation}
\tilde{c}_{\varrho _{1},\varrho _{2}}=\dsum \limits_{a,b,c}\Gamma
_{+c}^{+a_{1}}\times \left( t_{a_{1},\varrho _{1}}\otimes t_{a_{2},\varrho
_{2}}\right) \times \Gamma _{-c}^{-a_{2}}  \label{i2}
\end{equation}%
and where $\Gamma _{+c}^{+a}$ and $\Gamma _{-c}^{-a}$\ are as in eqs(\ref{gg}%
). To perform the explicit calculation of $\mathcal{I}_{1}$, we must know
the expression of the propagators $\left \langle \mathcal{B}_{\mu
}^{a}\left( X\right) \mathcal{B}_{\nu }^{b}\left( X^{\prime }\right)
\right
\rangle $; they are obtained form the propagators $G_{\rho \sigma
}^{cd}\left( X-X^{\prime }\right) =$ $\left \langle \mathcal{A}_{\rho
}^{c}\left( X\right) \mathcal{A}_{\sigma }^{d}\left( X^{\prime }\right)
\right \rangle $ of the CWY gauge field by using the relation $\mathcal{B}%
_{\mu }^{b}=\Gamma _{\mu a}^{\nu b}\mathcal{A}_{\nu }^{a}$. So, we have the
following relation 
\begin{equation}
\left \langle \mathcal{B}_{\mu }^{a}\left( X\right) \mathcal{B}_{\nu
}^{b}\left( X^{\prime }\right) \right \rangle =\Gamma _{\mu c}^{\rho
a}\times \Gamma _{\nu d}^{\sigma b}\times G_{\rho \sigma }^{cd}\left(
X-X^{\prime }\right)  \label{bc}
\end{equation}%
involving the two factor product $\Gamma _{\mu c}^{\rho a}\times \Gamma
_{\nu d}^{\sigma b}$. Recall that in the CWY theory, the free propagators $%
G_{\rho \sigma }^{cd}\left( X-X^{\prime }\right) $ are given by (\ref{pro})
that we re-express like 
\begin{equation}
G_{\rho \sigma }^{cd}\left( X_{1}-X_{2}\right) =\delta ^{cd}G_{\rho \sigma
}\left( X_{1}-X_{2}\right)
\end{equation}%
with%
\begin{equation}
G_{\rho \sigma }\left( X_{1}-X_{2}\right) =\frac{1}{2\pi }\varepsilon _{\rho
\sigma \tau }\eta ^{\tau \lambda }\frac{\partial \mathcal{R}_{12}}{\partial
X_{1}^{\lambda }}
\end{equation}%
and%
\begin{equation}
\mathcal{R}_{12}=\frac{1}{\left( X_{1}^{+}-X_{2}^{+}\right) \left(
X_{1}^{-}-X_{2}^{-}\right) +\left \vert \zeta _{1}-\zeta _{2}\right \vert
^{2}}
\end{equation}%
By substituting in (\ref{bc}), we get%
\begin{equation}
\left \langle \mathcal{B}_{\mu }^{a}\left( X\right) \mathcal{B}_{\nu
}^{b}\left( X^{\prime }\right) \right \rangle =\frac{1}{2}\Gamma _{\mu
c}^{\rho a}\times G_{\rho \sigma }\times \Gamma _{\nu c}^{\sigma b}
\label{bb}
\end{equation}%
where we have used eq(\ref{pro}) namely $G_{\rho \sigma }=\frac{1}{4\pi }%
\varepsilon _{\rho \sigma \tau }\eta ^{\tau \lambda }\frac{\partial \mathcal{%
R}}{\partial X^{\lambda }}$; and where summation on the repeated index c is
understood. Using these $\mathcal{B}$- propagators and following the method
of \cite{1F,1J} by choosing the first line operator as supported on the axis 
$X^{+}=x$ at $\zeta =\zeta _{1}$ and the second line operator as supported
on the axis $X^{-}=y$ at $\zeta =\zeta _{2}$, we can determine the explicit
value of then quantum contribution of the two line operators $K_{\zeta _{1}}$
and $K_{\zeta _{2}}$ exchanging one gluon. The calculations are quite
similar to the ones done in \cite{1F,1J} by using the generic $\mathcal{A}%
_{\mu }^{a}$'s; the novelty here is given by the fact that now we have a
restriction coming from the factors $\Gamma _{\mu a}^{\rho b}$ of (\ref{LC}%
). By setting $X=X_{1}-X_{2}$, the 2-form propagator $\mathcal{Q}^{ab}$
associated with $\left \langle \mathcal{B}_{\mu }^{a}\left( X_{1}\right) 
\mathcal{B}_{\nu }^{b}\left( X_{2}\right) \right \rangle $ is given by%
\begin{equation}
\mathcal{Q}^{ab}=\frac{1}{2}\left \langle \mathcal{B}_{\mu }^{a}\left(
X_{1}\right) \mathcal{B}_{\nu }^{b}\left( X_{2}\right) \right \rangle
dX^{\mu }\wedge dX^{\nu }
\end{equation}%
and, by using (\ref{bb}), it reads explicitly like%
\begin{equation}
\mathcal{Q}^{ab}=\frac{1}{2}\left( \Gamma _{\mu c}^{\rho a}\times G_{\rho
\sigma }\times \Gamma _{\nu c}^{\sigma b}\right) dX^{\mu }\wedge dX^{\nu }
\end{equation}%
This expression can be simplified by noticing from eq(\ref{gg}) that the
components of the tensor $\Gamma _{\mu c}^{\rho a}$ are proportional to
Kronecker $\eta _{\mu }^{\rho },$ it results therefore the following
relation between $\mathcal{Q}^{ab}=\mathcal{Q}^{ab}\left( X-X^{\prime
}\right) $\ and $\mathcal{P}=\mathcal{P}\left( X-X^{\prime }\right) $, 
\begin{equation}
\mathcal{Q}^{ab}=\Gamma _{+c}^{+a}\times \mathcal{P}\times \Gamma _{-c}^{-b}
\label{qp}
\end{equation}%
with $\mathcal{P}=\mathcal{P}_{+-}dX^{+}\wedge dX^{-}$ that reads explicitly
as 
\begin{equation}
\mathcal{P}=\frac{1}{2\pi }\frac{2\left( \bar{\zeta}_{1}-\bar{\zeta}%
_{2}\right) }{\left( x_{1}-x_{2}\right) ^{2}+\left( y_{1}-y_{2}\right)
^{2}+\left \vert \zeta _{1}-\zeta _{2}\right \vert ^{2}}dx\wedge dy
\end{equation}%
Comparing the above $\mathcal{Q}^{ab}$ quantity with the analogous $\mathcal{%
P}^{ab}=\delta ^{ab}\mathcal{P}$ given by (\ref{pp}), used in \textrm{\cite%
{1F}} for generic vector potentials $\mathcal{A}_{\mu }$, we end with the
following expression for the one- loop contribution 
\begin{equation}
\mathcal{I}_{1}=\hbar \left( t_{a,\varrho }\otimes t_{b,\varrho ^{\prime
}}\right) \dint \nolimits_{x,y}\mathcal{Q}^{ab}\left( x-x^{\prime
},y-y^{\prime };\zeta _{1}-\zeta _{2},\bar{\zeta}_{1}-\bar{\zeta}_{2}\right)
\end{equation}%
By substituting $\mathcal{Q}^{ab}$ by its expression (\ref{qp}), we also have%
\begin{equation}
\mathcal{I}_{1}=\hbar _{\text{ }}\tilde{c}_{\varrho ,\varrho ^{\prime
}}\dint \nolimits_{x,y}\mathcal{P}\left( x-x^{\prime },y-y^{\prime };\zeta
_{1}-\zeta _{2},\bar{\zeta}_{1}-\bar{\zeta}_{2}\right)
\end{equation}%
with $c_{\varrho ,\varrho ^{\prime }}$ reading as 
\begin{equation}
\tilde{c}_{\varrho ,\varrho ^{\prime }}=\dsum \limits_{a,b,c}\Gamma
_{+c}^{+a}\times \left( t_{a,\varrho }\otimes t_{b,\varrho ^{\prime
}}\right) \times \Gamma _{-c}^{-b}
\end{equation}%
and the two integrations over $\mathcal{P}$ given by $\frac{1}{\zeta
_{1}-\zeta _{2}}$ in agreement with the scaling dimension and the invariance
of the 1-form $\omega _{1}=d\zeta $ in the field action $\mathcal{S}_{cwy}$
under global translation $\zeta ^{\prime }=\zeta +cte$. The construction we
have done above for the sl$\left( 2\right) $ case of the Liouville equation
extends straightforwardly to finite dimensional Lie algebras of 2d Toda QFT$%
_{2}$; \textrm{see appendix section}.

\section{Conclusion}

By using the Lax formalism of 2d integrable models, we have studied in this
paper the embedding of finite Toda QFT$_{2}$'s into the
Costello-Witten-Yamazaki theory by focussing on the first term in this
family of 2d integrable models namely the Liouville model. After reviewing
briefly some useful aspects of Liouville theory and properties of the Lax
method of 2d integrable systems, we have shown how the two Lax operators $%
L_{\pm }$ of Liouville equation can be derived from the 4d CWY gauge
connection $\mathcal{A}=t_{a}\mathcal{A}_{\mu }^{a}dx^{\mu }$ with SL$\left(
2\right) $ gauge symmetry. This has been done by imposing appropriate
constraints on some $\mathcal{A}_{\mu }^{a}$ components of the non abelian
gauge potential $\mathcal{A}_{\mu }=t_{a}\mathcal{A}_{\mu }^{a}$ and
interpreted in terms of turning off propagation of $\mathcal{A}_{\mu }^{a}$
in some directions of the gauge bundle. By using results from \cite{1F,1J}
regarding the observables into CWY theory, we have constructed quantum line
operators that are associated with the Lax pair describing the Liouville
field equation; this construction extends straightforwardly to the family of
2d Toda QFT$_{2}$'s based on finite Lie algebras $\mathcal{G}_{c}$. As an
illustration of quantum excitations of these lines, we have also computed
the one loop contribution of two interesting line operators exchanging one
gauge boson and shown how the quasi-classical solution of the R-matrix gets
modified compared to the result of \cite{1F}. We suspect that the method
developed in this study may be applied to a large class of 2d integrable
systems especially to those systems having a Lax pair formulation; the ones
considered in this paper concern the class of 2d Toda QFT$_{2}$ based on
finite dimensional Lie algebra. It would be interesting to extend this
construction to other families like for instance the affine 2d Toda QFT$_{2}$%
's containing the sinh-Gordon model. From the analysis given in this study,
we expect that these kinds of integrable systems correspond as well to
particular orientations of the non abelian gauge potential $\mathcal{A}_{\mu
}=t_{a}\mathcal{A}_{\mu }^{a}$ in the gauge bundle; these directions are
given by the $\Gamma _{\mu a}^{\rho b}$ of eq(\ref{LC}) and so this tensor
may be also interpreted as an object that characterise the 2d integrable
models.

\  \  \  \  \newline
We end this conclusion by noticing that the constraint eqs(\ref{523}) need
to be deepen much more as they constitute a key step towards the embedding
of Lax pairs into the CWY theory. A rough way to approach these gauge
conditions is by using standard methods for dealing with constrained systems
like the Lagrange multipliers method; then look how CWY theory and its
observables may be modified. In this regards, we would like to highlight%
\textrm{\ }some basic points that still need a refinement before talking
about a fully consistent embedding of Lax pairs of 2d integrable models into
CWY theory. For example, in our present study we have not been able to work
out a simple mechanism that can naturally give rise to the constraints eqs(%
\ref{523}) by starting from CWY theory. Recall that in quantum field theory,
fields are to be integrated over all allowed configurations in the path
integral, and one cannot arbitrarily impose constraints by hand as we have
done where we have used known results on Liouville theory and its extension.
Clearly, additional ingredients are needed to realize (\ref{523}) in a
physical manner; perhaps the introduction of a surface defect may be a good
candidate to explore. Moreover, even if one could realize these constraints
in some simple way, a question still remain to answer as it is not clear how
the coupling constant $\kappa $ of Liouville theory can be determined from
the CWY equations; the coupling is given by the product $\kappa =\alpha
\beta $ of the overall factors appearing in $A_{+}^{-}=-\alpha $ and $%
A_{-}^{+}=\beta e^{2\phi }$ as shown by (4.24)-(4.25). Another interesting
point concerns as well the explicit study of gauge fixing and gauge symmetry
reduction. Recall that the CWY theory has a gauge symmetry which one has to
take care of when quantizing the theory; in our study the gauge symmetry is
disturbed by the constraints \textrm{(\ref{523}); }a separate treatment is
therefore required. Progress in these directions will be reported in a
future occasion.

\section{Appendix: More on Toda field models embedding}

In this appendix, we give further details concerning the embedding of
generic 2d finite Toda field models in the Costello-Witten-Yamazaki theory.
These Toda models are based on finite dimensional Lie algebras with rank r%
\TEXTsymbol{>}1; they extend the Liouville equation relying on the sl$\left(
2\right) $ Lie algebra treated as an example in the core of the paper. After
describing some properties of this family of 2d integrable systems, we build
the Lax pair L$_{\pm }$ for these field models and comment on some of their
useful properties. Then, we describe the main lines of the embedding of this
special family of conformal models into the 4d topological CWY theory. We
end up this appendix by making comments and giving some remarkable aspects
of this construction; in particular the issue concerning conformal
invariance along the exotic $\zeta $-dimension.

\subsection{Finite Toda models}

Two dimensional finite Toda field theory (FTFT) are integrable systems
extending the 2d Liouville field equation introduced in section 3. This is a
family of 2d solvable field models in the sens of Lax formulation; it has
rich symmetries including the infinite dimensional 2d conformal symmetry of
the Liouville theory given by Eqs(\ref{c1}-\ref{ct}); but it also has other
infinite symmetries beyond 2d conformal invariance known as W- invariance 
\textrm{\cite{1W,2W,3W,4W,5W,6W,7W,8W,9W}. It is generally admitted that it
is these infinite symmetries that underly the integrability of this class of
highly non linear 2d field theory}. Finite Toda field models are classified
by Cartan classification of finite dimensional Lie algebras $\mathcal{G}$
including the simply laced $ADE$ and the non simply laced BCFG algebras. The
Dynkin diagrams of these algebras are as depicted by the Figure \textbf{\ref%
{abc}}. 
\begin{figure}[tbph]
\begin{center}
\hspace{0cm} \includegraphics[width=14cm]{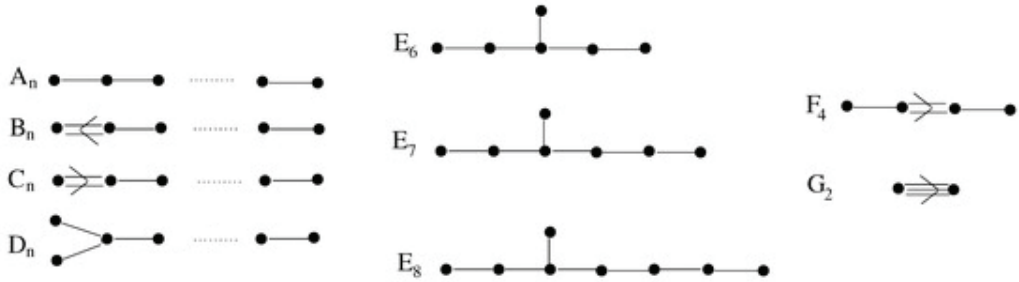}
\end{center}
\par
\vspace{-0.5 cm}
\caption{Dynkin diagrams of finite dimensional Lie algebras with rank r=n
given by the number of nodes. The diagram of A$_{1}$ represented by one node
is the simplest one.}
\label{abc}
\end{figure}

\  \  \newline
The field action $\mathcal{S}=\int_{\Sigma }d^{2}\xi \mathcal{L}$ of these
2d conformal FTFT is given by Eq(\ref{tod}) whose variation leads to r field
equations of motion given by%
\begin{equation}
\partial _{+}\partial _{-}\phi _{i}+\kappa _{i}\exp \left(
\sum_{j=1}^{r}C_{ij}\phi _{j}\right) =0  \tag{A1}  \label{toda}
\end{equation}%
with $C_{ij}$ standing for the Cartan matrix which reads in terms of the
simple roots of the Lie algebra as $\frac{2}{\mathbf{\alpha }_{i}\mathbf{%
.\alpha }_{i}}\mathbf{\alpha }_{i}\mathbf{.\alpha }_{j}$. For explicit
calculations, we will focuss in what follows on the $A_{r}$ subfamily where $%
\mathbf{\alpha }_{i}\mathbf{.\alpha }_{i}=2$ leading to a symmetric Cartan
matrix; however the analysis which will be developed below applies as well
for the other subfamilies of finite Toda field models. To begin notice that
for the leading rank one example $A_{1}$, the above equation reduces to the
Liouville equation (\ref{1f}) with Liouville field given by $\phi _{1}\left(
\xi ,\bar{\xi}\right) $. For the rank r=2 algebra $A_{2},$ with Cartan
matrix $C_{ij}$ given by the intersection matrix of the two possible simple
roots $\mathbf{\alpha }_{1}$ and $\mathbf{\alpha }_{2}$ namely%
\begin{equation}
C_{ij}=\left( 
\begin{array}{cc}
2 & -1 \\ 
-1 & 2%
\end{array}%
\right) \qquad ,\qquad \left( C_{ij}\right) ^{-1}=\frac{1}{3}\left( 
\begin{array}{cc}
2 & 1 \\ 
1 & 2%
\end{array}%
\right) ,  \tag{A2}
\end{equation}%
the $A_{2}$- Toda field equations resulting from (\ref{toda}) are given by
the two following coupled partial differential equations%
\begin{equation}
\begin{tabular}{lll}
$\partial _{+}\partial _{-}\phi _{1}+\kappa _{1}e^{2\phi _{1}-\phi _{2}}$ & $%
=$ & $0$ \\ 
$\partial _{+}\partial _{-}\phi _{2}+\kappa _{2}e^{2\phi _{2}-\phi _{1}}$ & $%
=$ & $0$%
\end{tabular}
\tag{A3}  \label{12}
\end{equation}%
where $\kappa _{1}$ and $\kappa _{2}$ are two real coupling parameters that
we assume positive definite;\textrm{\ they are the homologue of the
Liouville coupling constant }$\kappa =\alpha \beta $\textrm{\ with }$\alpha $%
\textrm{\ and }$\beta $ \textrm{interpreted in this paper as a kind of field
VEVs as exhibited by Eqs(\ref{23})}. By multiplying these equations
respectively by the diagonal $h_{1}$ and $h_{2}$ Cartan generators of $A_{2}$%
, one can rewrite them as follows%
\begin{equation}
\begin{tabular}{lll}
$\left( \partial _{+}\partial _{-}\phi _{1}+\kappa _{1}e^{2\phi _{1}-\phi
_{2}}\right) h_{1}$ & $=$ & $0$ \\ 
$\left( \partial _{+}\partial _{-}\phi _{2}+\kappa _{2}e^{2\phi _{2}-\phi
_{1}}\right) h_{2}$ & $=$ & $0$%
\end{tabular}
\tag{A4}  \label{13}
\end{equation}%
or equivalently by adding them like 
\begin{equation}
\left( \partial _{+}\partial _{-}\phi _{1}+\kappa _{1}e^{2\phi _{1}-\phi
_{2}}\right) h_{1}+\left( \partial _{+}\partial _{-}\phi _{2}+\kappa
_{2}e^{2\phi _{2}-\phi _{1}}\right) h_{2}=0  \tag{A5}  \label{130}
\end{equation}%
With this trick, Eqs(\ref{12}) appear as nothing but projections of the
matrix Eqs (\ref{13}) or their linear combination (\ref{130}) along the two
Cartan directions of A$_{2}$. In general, the finite A$_{r}$- Toda equations
(\ref{toda}) ---and their homologue for the other Lie algebras $\mathcal{G}$%
--- can be also viewed as nothing else but the projections of the diagonal
matrix equation%
\begin{equation}
\sum_{i=1}^{r}\left( \partial _{+}\partial _{-}\phi _{l}+\kappa
_{i}e^{C_{ij}\phi _{j}}\right) h_{l}=0  \tag{A6}  \label{15}
\end{equation}%
along the $h_{i}$-th directions in the Cartan torus of the Lie algebra. In
this set up, one can use Lax formalism to linearise this matrix equation by
working out a Lax pair L$_{\pm }$ in terms of the Toda fields $\phi _{l}$
and the generators of the A$_{r}$ Lie algebra. Generally speaking, field
realisations of this L$_{\pm }$ pair can be motivated by using dimensional
scaling properties and symmetry arguments; it has the following typical form 
\begin{equation}
\begin{tabular}{lll}
$L_{+}$ & $=$ & $h_{i}\partial _{+}\phi _{i}-\psi _{i}E_{i}^{+}$ \\ 
$L_{-}$ & $=$ & $h_{i}\partial _{-}\omega _{i}-\chi _{i}E_{i}^{-}$%
\end{tabular}
\tag{A7}  \label{mr}
\end{equation}%
where sigma- summation over the repeated index i is understood. In these
expressions, the r triplets $\left( h_{i},E_{i}^{\pm }\right) $ are the
Chevalley generators of the Lie algebra obeying the commutation relations%
\begin{equation}
\begin{tabular}{lll}
$\left[ h_{i},E_{j}^{\pm }\right] $ & $=$ & $\pm C_{ij}E_{j}^{\pm }$ \\ 
$\left[ E_{i}^{+},E_{j}^{-}\right] $ & $=$ & $\delta _{ij}h_{i}$%
\end{tabular}
\tag{A8}  \label{16}
\end{equation}%
while the coefficients $\phi _{i},\psi _{i},\omega _{i}$ and $\chi _{i}$ are
2d fields whose relationships are obtained by partial solving of the matrix
Lax equation 
\begin{equation}
\partial _{+}L_{-}-\partial _{-}L_{+}+\left[ L_{+},L_{-}\right] =0  \tag{A9}
\label{lax}
\end{equation}%
By substituting (\ref{mr}), this matrix equation can be cast into three sets
of \textquotedblleft smaller\textquotedblright \ matrix equations in one to
one with the $h_{i}$- and $E_{i}^{\pm }$- Chevalley directions as follows%
\begin{equation}
\begin{tabular}{lll}
$\left[ \partial _{+}\partial _{-}\left( \phi _{i}-\omega _{i}\right) -\psi
_{i}\chi _{i}\right] h_{i}$ & $=$ & $0$ \\ 
$\  \left[ \partial _{-}\psi _{l}+\psi _{l}\partial _{-}\left( \omega
_{i}C_{il}\right) \right] E_{l}^{+}$ & $=$ & $0$ \\ 
$\  \left[ \partial _{+}\chi _{l}-\chi _{l}\partial _{+}\left( \phi
_{i}C_{il}\right) \right] E_{l}^{-}$ & $=$ & $0$%
\end{tabular}
\tag{A10}
\end{equation}%
The two last equations should be understood as constraint equations while
the first relation gives the Toda field equations we are interested in here.
Notice that the $h_{i}$- equation above suggests also that it is convenient
to set $\omega _{i}=0$ as only the combination $\phi _{i}-\omega _{i}$ which
propagate; this choice allows to reduce the second equation down to $%
\partial _{-}\psi _{l}=0$ restricting thus $\psi $ to a holomorphic function 
$\psi \left( X^{+}\right) $. \newline
In conclusion to this subsection and as a result for FTFT, an explicit
realisation of the $L_{\pm }$ Lax pair in terms of the Toda fields and the
Chevalley generators reads as follows 
\begin{equation}
\begin{tabular}{lll}
$L_{+}$ & $=$ & $h_{i}\partial _{+}\phi _{i}-\epsilon _{i}E_{i}^{+}$ \\ 
$L_{-}$ & $=$ & $\beta e^{C_{ij}\phi _{j}}E_{i}^{-}$%
\end{tabular}
\tag{A11}  \label{LL}
\end{equation}%
where $\epsilon _{i}$ and $\beta $ are constant parameters defining the Toda
coupling constants $\kappa _{i}=\beta \epsilon _{i}$. Notice that one may
also take $r$ parameters $\beta _{i}$ instead of one parameter $\beta $; in
this case $L_{-}$ is realised by the sum over $\beta _{i}e^{C_{ij}\phi
_{j}}E_{i}^{-}$ and then $\kappa _{i}=\beta _{i}\epsilon _{i}$; this is not
important at this level; but might be relevant for other issues. Notice also
that for the case r=1, one recovers Eq(\ref{4}) of the Liouville theory.

\subsection{Embedding FTFT in CWY theory}

In this subsection, we describe the main lines for the embedding of finite
Toda field models described above in the CWY construction. We first
introduce the generalised pairs $\left( \mathcal{L}_{+},\mathcal{L}%
_{-}\right) $ and $\left( \mathcal{L}_{\bar{\zeta}},\mathcal{L}_{\pm
}\right) $; then we draw the steps for the embedding of the FTFT in the CWY
theory. \textrm{We close this study by making comments and giving some
remarks.}

\textrm{\ } \newline
Given the Lax pair L$_{\pm }$ realisation (\ref{LL}) and the associated Lax
equation (\ref{lax}), one can extend straightforwardly the embedding study
done for the Liouville equation in section 5 to the finite Toda field
models. To that purpose, it is interesting to start from some useful
observations. First, think of the pair $L_{\pm }$, living in 2d space
dimension, as a restriction $\left. \mathcal{L}_{\pm }\right \vert $ of a
generalised pair $\left( \mathcal{L}_{+},\mathcal{L}_{-}\right) $ living in
the 4d CWY space--- parameterised by $\left( X^{\pm },\zeta ,\bar{\zeta}%
\right) $ with $X^{\pm }$ denoted as well by $\left( \xi ,\bar{\xi}\right) $%
--- and satisfying 
\begin{equation}
\partial _{+}\mathcal{L}_{-}-\partial _{-}\mathcal{L}_{+}+\left[ \mathcal{L}%
_{+},\mathcal{L}_{-}\right] =0  \tag{A12}  \label{21}
\end{equation}%
In other words, we have 
\begin{equation}
L_{\pm }\left( X\right) =\left. \mathcal{L}_{\pm }\left( X;\zeta ,\bar{\zeta}%
\right) \right \vert _{\zeta =\bar{\zeta}=0}  \tag{A13}  \label{res}
\end{equation}%
But because of the fact that the space dimension in CWY theory is bigger
than the space dimension in standard FTFT, one has, in addition to $\left( 
\mathcal{L}_{+},\mathcal{L}_{-}\right) ,$ two extra Lax-like pairs $\left( 
\mathcal{L}_{\bar{\zeta}},\mathcal{L}_{+}\right) $ and $\left( \mathcal{L}_{%
\bar{\zeta}},\mathcal{L}_{-}\right) $ with coordinate dependence as above;
see also (\ref{lp}). These two extra pairs satisfy as well the following
Lax-type equations%
\begin{equation}
\begin{tabular}{lll}
$\partial _{\bar{\zeta}}\mathcal{L}_{+}-\partial _{+}\mathcal{L}_{\bar{\zeta}%
}+\left[ \mathcal{L}_{\bar{\zeta}},\mathcal{L}_{+}\right] $ & $=$ & $0$ \\ 
$\partial _{\bar{\zeta}}\mathcal{L}_{-}-\partial _{-}\mathcal{L}_{\bar{\zeta}%
}+\left[ \mathcal{L}_{\bar{\zeta}},\mathcal{L}_{-}\right] $ & $=$ & $0$%
\end{tabular}
\tag{A14}  \label{22}
\end{equation}%
Notice that the missing pairs involving the operator $\mathcal{L}_{\zeta }$
namely $\left( \mathcal{L}_{\zeta },\mathcal{L}_{\pm }\right) $ and $\left( 
\mathcal{L}_{\zeta },\mathcal{L}_{\bar{\zeta}}\right) $ are forbidden in the
CWY construction as there is no propagation along the $\zeta $- direction.

\  \  \  \newline
To embed the finite Toda field models in CWY theory, we use (\ref{res}) and
the field realisation (\ref{LL}) teaching us that $\mathcal{L}_{\pm }$
should have a quite similar Lie algebraic structure compared to $L_{\pm }$.
By following the method used in previous subsection and taking advantage of
Eqs(\ref{40}) describing the embedding Liouville equation; it is not
difficult to check that the following field realisation%
\begin{equation}
\begin{tabular}{lll}
$\mathcal{L}_{+}$ & $=$ & $\frac{\partial \Phi _{i}}{\partial X^{+}}%
h_{i}-N_{i}E_{i}^{+}$ \\ 
$\mathcal{L}_{-}$ & $=$ & $\Psi e^{C_{ij}\Phi _{j}}E_{i}^{-}$ \\ 
$\mathcal{L}_{\bar{\zeta}}$ & $=$ & $-\frac{1}{2}\frac{\partial \log N_{i}}{%
\partial \bar{\zeta}}h_{i}+\Gamma e^{C_{ij}\Phi _{j}}E_{i}^{-}$%
\end{tabular}
\tag{A15}  \label{l}
\end{equation}%
is indeed a solution of the generalised Lax equations (\ref{21}-\ref{22}).
In this representation, we have 2r+2 fields namely $\Phi _{i},N_{i},\Psi $
and $\Gamma $; they are the homologue of the 4 fields appearing in (\ref{40}%
) concerning A$_{1}$. Notice that this realisation has two remarkable
properties that we want to comment on before proceeding. First, as expected
the $\mathcal{L}_{+}$ and $\mathcal{L}_{-}$ are not related to each other by
adjoint conjugation. Moreover, $\mathcal{L}_{+}$ carries the propagation of
the $\Phi _{i}$'s along the $X^{+}$-direction; while $\mathcal{L}_{-}$
carries interactions among the r fields. Regarding $\mathcal{L}_{\bar{\zeta}%
} $, it carries propagation in the $\bar{\zeta}$- direction and also
interactions. The lack of a propagation in $X^{-}$-direction is only
apparent; it is manifested through the second relation of Eqs(\ref{22}).
Second, the realisation (\ref{l}) involves only the Chevalley generators $%
h_{i}$, $E_{i}^{\pm }$ of the A$_{r}$ Lie algebra: the $E_{\mathbf{\gamma }%
}^{\pm }$ operators associated with non simple roots $\mathbf{\gamma }$ of
the underlying Lie algebra do not appear in our embedding. Missing terms
will be discussed later on; see after eq(\ref{aam}).

\  \  \  \newline
The next step to do is to think on Eqs(\ref{21}) and (\ref{22}) as the
vanishing condition of the\ CWY curvature of a constrained gauge potential $%
\mathcal{A}_{\mu }$; that is as corresponding to the zero of%
\begin{equation}
\mathcal{F}_{\mu \nu }=\partial _{\mu }\mathcal{A}_{\nu }-\partial _{\nu }%
\mathcal{A}_{\mu }+\left[ \mathcal{A}_{\mu },\mathcal{A}_{\nu }\right] 
\tag{A16}
\end{equation}%
with $\mathcal{L}_{\mu }=\left. \mathcal{A}_{\mu }\right \vert $ with $\mu
=\pm ,\bar{\zeta}$. By following the same method done in subsection 5.2
including the remarks; one can derive the constraints that have to be
imposed on the CWY gauge potential $\mathcal{A}_{\mu }=t_{a}\mathcal{A}_{\mu
}^{a}$ to recover the classical finite Toda field equations. These
constraint eqs can be worked out by using the Cartan basis $\left(
h_{i},E_{\pm \alpha }\right) $ of the Lie algebra and expanding the non
abelian gauge potential matrix as follows%
\begin{equation}
\mathcal{A}_{\mu }=\sum_{i=1}^{r}h_{i}\mathcal{A}_{\mu }^{i}+\sum_{\func{%
positive}\text{ roots }\gamma }\left( E_{+\mathbf{\gamma }}\mathcal{A}_{\mu
}^{+\mathbf{\gamma }}+E_{-\mathbf{\gamma }}\mathcal{A}_{\mu }^{-\mathbf{%
\gamma }}\right)  \tag{A17}  \label{aam}
\end{equation}%
Comparing this expansion with the realisation (\ref{l}), one learns that $%
\mathcal{A}_{\mu }^{a}$ has to obey two kinds of constraint relations: the
first set of constraints is associated with the Lie algebraic index $a$; and
the second one with the space index $\mu =\pm ,\bar{\zeta}$. For the first
set, the constraints are given by the vanishing of $\mathcal{A}_{\mu }^{\pm 
\mathbf{\tau }}=tr\left( E_{\pm \mathbf{\tau }}\mathcal{A}_{\mu }\right) =0$
with $\mathbf{\tau }$ referring to non simple positive roots; i.e: $\mathbf{%
\tau }\neq \mathbf{\alpha }_{i}$. This is because the expansion of (\ref{l})
involves only Chevalley generators $h_{i}$, $E_{\pm \mathbf{\alpha }%
_{i}}=E_{i}^{\pm }$. The second set of constraints can be learnt directly
from the realisation (\ref{l}) by rewriting it as follows 
\begin{equation}
\begin{tabular}{lllll}
$\mathcal{L}_{+}$ & $=$ & $\frac{\partial \Phi _{i}}{\partial X^{+}}h_{i}$ & 
$-N_{i}E_{i}^{+}$ & $+0_{i}E_{i}^{-}$ \\ 
$\mathcal{L}_{-}$ & $=$ & $0_{i}$ $h_{i}$ & $+_{\text{ }}0_{i}$ $E_{i}^{+}$
& $+\Psi e^{C_{ij}\Phi _{j}}E_{i}^{-}$ \\ 
$\mathcal{L}_{\bar{\zeta}}$ & $=$ & $\frac{\partial \log N_{i}}{-2\partial 
\bar{\zeta}}h_{i}$ & $+_{\text{ }}0_{i}$ $E_{i}^{+}$ & $+\Gamma
e^{C_{ij}\Phi _{j}}E_{i}^{-}$%
\end{tabular}
\tag{A18}
\end{equation}%
From these generalised $\mathcal{L}_{\pm ,\bar{\zeta}}$ operators, one can
build the 1-form $\mathcal{L}=dX^{+}\mathcal{L}_{+}+dX^{-}\mathcal{L}_{-}+d%
\bar{\zeta}\mathcal{L}_{\bar{\zeta}}$ that can be imagined as a particular
1-form CWY gauge connection $\mathcal{A}=dX^{+}\mathcal{A}_{+}+dX^{-}%
\mathcal{A}_{-}+d\bar{\zeta}\mathcal{A}_{\bar{\zeta}}$. Equating $\mathcal{A}%
_{\pm },\mathcal{A}_{\bar{\zeta}}$ with the generalised Lax operators $%
\mathcal{L}_{\pm },\mathcal{L}_{\bar{\zeta}}$, we end up with the following
constraint relations 
\begin{equation}
\mathcal{A}_{-}^{i}=0\quad ,\quad \mathcal{A}_{+}^{+\alpha _{i}}=0\quad
,\quad \mathcal{A}_{-}^{-\alpha _{i}}=0\quad ,\quad \mathcal{A}_{\bar{\zeta}%
}^{-\alpha _{i}}=0  \tag{A19}
\end{equation}%
where the $\alpha _{i}$'s are the r simple roots of the Lie algebra. The
remaining step of the calculations, including the building of the $\mathcal{B%
}_{\mu }=\Gamma _{\mu a}^{\rho }\mathcal{A}_{\rho }^{a}$ potential as well
as the computation of the corresponding one loop quantum effect is achieved
by following similar lines used for sl$\left( 2\right) $. We omit these
details here; they may be obtained from the sl$\left( 2\right) $ ones by
quasi-direct extension; for instance the construction of the $\Gamma _{\mu
a}^{\rho }$'s can be deduced from (\ref{gg}); the term $\mathbf{\Gamma }%
_{-a}^{\rho }$ is given by $\eta _{-}^{\rho }E_{i}^{-}Tr\left(
E_{i}^{+}t_{a}\right) $.

\  \  \  \newline
We end this appendix by two more comments and some remarks; the first
comment concerns bi-holomorphic symmetries in the 4d space $\mathbb{R}%
^{2}\times \mathcal{C}$ which can be thought of as $\mathbb{C}\times \mathbb{%
C}$; and the second one regards the dependence of the \textrm{Toda field}
into the exotic variables $\zeta ,\bar{\zeta}$. Focussing on the leading A$%
_{1}$ model and too particularly on Eqs(\ref{ll1}-\ref{ll2}) that we rewrite
like%
\begin{eqnarray}
\partial _{-}\partial _{+}\Phi +K_{-+}e^{2\Phi } &=&0\qquad ,\qquad
K_{-+}=\Psi _{-}N_{+}  \TCItag{A20}  \label{pn1} \\
\partial _{\bar{\zeta}}\partial _{+}\Phi +K_{\bar{\zeta}+}e^{2\Phi }
&=&0\qquad ,\qquad K_{\bar{\zeta}+}=\Gamma _{\bar{\zeta}}N_{+}  \TCItag{A21}
\label{pn2}
\end{eqnarray}%
where $\partial _{\pm }=\frac{\partial }{\partial X^{\pm }}$ and $\partial _{%
\bar{\zeta}}=\frac{\partial }{\partial \bar{\zeta}}$,\ one learns that the
4d generalised Liouville field $\Phi \left( X^{\pm },\zeta ,\bar{\zeta}%
\right) $ living in the CWY space has richer symmetries compared to the
standard 2d Liouville field $\phi \left( X^{\pm }\right) $. Indeed, because
of lack of propagation along the $\zeta $-direction in the field action $%
\mathcal{S}_{cyw}$ given by Eq(\ref{1}), it is clear that the above Eqs(\ref%
{pn1}-\ref{pn2}) have propagations in the three $X^{\pm },\bar{\zeta}$
directions; they have as well infinite dimensional symmetries given by (\ref%
{t1}-\ref{a}) that we present as follows 
\begin{equation}
\begin{tabular}{lll}
$\Phi ^{\prime }$ & $=$ & $\Phi +\frac{1}{2}\mathbf{k}\left( \zeta \right) -%
\frac{1}{2}\mathbf{f}\left( \zeta \right) $ \\ 
$N_{+}^{\prime }$ & $=$ & $e^{-\mathbf{k}\left( \zeta \right) }N_{+}$ \\ 
$\Psi _{-}^{\prime }$ & $=$ & $e^{\mathbf{f}\left( \zeta \right) }\Psi _{-}$
\\ 
$\Gamma _{\bar{\zeta}}^{\prime }$ & $=$ & $e^{\mathbf{f}\left( \zeta \right)
}\times \Gamma _{\bar{\zeta}}$%
\end{tabular}
\tag{A22}  \label{tc}
\end{equation}%
where $\mathbf{k}\left( \zeta \right) $ and $\mathbf{f}\left( \zeta \right) $
are arbitrary holomorphic functions in the $\zeta $\ variable. By comparing
the above transformations of $\Phi $ with the conformal transformation (\ref%
{ct}) in the 2d space parameterised by X$^{\pm }$, which holds as well for (%
\ref{pn1}-\ref{pn2}), it results that the generalised Liouville- like
equations have a bi-holomorphic invariance; that is two kinds of 2d
conformal invariance: A first 2d conformal symmetry given by the holomorphic
coordinate change (\ref{ct}) in the $\mathbb{R}^{2}\sim \mathbb{C}$ part of
the CWY space $\mathbb{R}^{2}\times \mathcal{C}$ with $N_{+}$ and $\Psi _{-}$
transforming respectively like $\frac{\partial }{\partial X^{+}}$ and $\frac{%
\partial }{\partial X^{-}}$. A second 2d conformal invariance given by (\ref%
{tc}); this symmetry can be also viewed at the level of the CWY\ action $%
\mathcal{S}_{cyw}$ showing that $\boldsymbol{\Omega }_{3}$ should be
invariant under the holomorphic coordinate change $\zeta \rightarrow f\left(
\zeta \right) $ which leaves the 1-form $\mathbf{\omega }_{1}=d\zeta \omega
_{\zeta }$ unchanged. \newline
Concerning, the second comment, notice that multiplying Eq(\ref{pn1}) from
left by $X^{-}$, and doing the same thing for Eq(\ref{pn2}) but multiplying
by $\bar{\zeta}$, then subtracting the two resulting relations, we get%
\begin{equation}
\frac{\partial }{\partial X^{+}}\left( X^{-}\frac{\partial \Phi }{\partial
X^{-}}-\bar{\zeta}\frac{\partial \Phi }{\partial \bar{\zeta}}\right) =0 
\tag{A23}
\end{equation}%
provided $X^{-}\Psi _{-}=\bar{\zeta}\Gamma _{\bar{\zeta}}$. By demanding the
relative quantity $X^{-}\frac{\partial \Phi }{\partial X^{-}}-\bar{\zeta}%
\frac{\partial \Phi }{\partial \bar{\zeta}}$ to be independent of $X^{+}$,
the two equations (\ref{pn1}) and (\ref{pn2}) merge into one equation.
Moreover, by using Eq(\ref{a}) and substituting the expressions of $\Psi
_{-} $ and $\Gamma _{\bar{\zeta}}$ back into $X^{-}\Psi _{-}=\bar{\zeta}%
\Gamma _{\bar{\zeta}}$, we end up with the result that the field equation of
motion of the\ generalized Liouville $\Phi $ lives precisely on the line $%
\bar{\zeta}=\frac{\beta }{\gamma }X^{-}$ in the CWY space.

\begin{acknowledgement}
: This work is supported by project "Topological phase of matter" Hassan II
Academy of Science and Technology.
\end{acknowledgement}

\end{document}